# CLIMATE SIMULATIONS OF EARLY MARS WITH ESTIMATED PRECIPITATION, RUNOFF, AND EROSION RATES

Ramses M. Ramirez[1,2], Robert A. Craddock[3] and Tomohiro Usui[1,4]

[1]Earth-Life Science Institute, Tokyo Institute of Technology, Tokyo, Japan
[2]Space Science Institute, Boulder, Co, USA
[3]Center for Earth and Planetary Studies, National Air and Space Museum, Smithsonian Institution, Washington D.C., USA
[4]Institute of Space and Astronautical Science (ISAS), Japan Aerospace Exploration Agency, Tokyo, Japan

Corresponding author: Ramses Ramirez (rramirez@elsi.jp)

**Key Points:**

- Mean surface temperatures near or slightly above the freezing point of water produce climates that could have carved the valleys

- Such scenarios are consistent with a relatively large northern lowlands ocean

- Our results support a warm and semi-arid climate for early Mars and can explain many geologic observations

**Abstract**

The debate over the early Martian climate is among the most intriguing in planetary science. Although the geologic evidence generally supports a warmer and wetter climate, climate models have had difficulty simulating such a scenario, leading some to suggest that the observed fluvial geology (e.g. valley networks, modified landscapes) on the Martian surface, could have formed in a cold climate instead. However, as we have originally predicted using a single-column radiative-convective climate model [*Ramirez et al*. 2014a], warming from $CO_2$-$H_2$ collision-induced absorption (CIA) on a volcanically active early Mars could have raised mean surface temperatures above the freezing point, with later calculations showing that this is achievable with hydrogen concentrations as low as ~1%. Nevertheless, these predictions should be tested against more complex models. Here, we use an advanced energy balance model that includes a northern lowlands ocean to show that mean surface temperatures near or slightly above the freezing point of water were necessary to carve the valley networks. Our scenario is consistent with a relatively large ocean as has been suggested. Valley network distributions would have been global prior to subsequent removal processes. At lower mean surface temperatures and smaller ocean sizes, precipitation and surface erosion efficiency diminish. The warm period may have been ~< $10^7$ years, perhaps suggesting that episodic warming mechanisms were not needed. Atmospheric collapse and permanently glaciated conditions occur once surface ice coverage exceeds a threshold depending on CIA assumptions. Our results support an early warm and semi-arid climate consistent with many geologic observations.

**Plain Language Summary**
Mars today is a dry and cold planet that looks quite "Moon-like." However, ancient landscapes on Mars depict a world that may have been much more Earth-like billions of years ago. The surface is filled with ancient riverbeds, deltas, and even features that look a lot like ancient ocean shorelines. Some scientists think that such features suggest a warmer and wetter early planet whereas others believe that these features had formed when the planet was not too different from today.

Here, we test the idea that a warm early Mars could have had a large northern ocean sustained by a thick carbon-dioxide and hydrogen atmosphere. We first confirm previous studies that showed that such an atmosphere could have made Mars warm. We then demonstrate that a relatively large ocean would have been required to form the water features we see on the ancient surface. It may have taken as little as 10 million years or less to form them. In contrast, cold and icy climates, even if they were sporadically warmed from time to time, may not have produced enough precipitation to form these features. We confirm that rain would have been the dominant precipitation on a warm early Mars.

# 1 Introduction

The climate of early Mars is one of the most intriguing problems in planetary science. Although Mars is mostly dry and possesses a tenuous atmosphere today, this was not the case ~ 3.8 – 4 billion years ago when the geologic evidence reveals a landscape filled with many fluvial features, including deltas, alluvial fans, ancient shorelines, modified craters, and of course, the enigmatic valley networks [e.g., *Masursky et al.* 1973; *Parker et al.* 1993*; Craddock and Howard* 2002; *Howard et al.* 2005; *Fasset and Head* 2008; *Irwin et al.* 2011]. This geologic evidence is strongly indicative of a once warmer and wetter climate than is the case today [e.g., *Craddock and Howard* 2002; *Ramirez and Craddock* 2018].

In spite of the geologic observations, however, both 1-D and 3-D climate models have struggled to simulate warm conditions for early Mars. Although it is predicted that Mars may have had a dense $CO_2$ atmosphere in the past [e.g., *Pollack et al.* 1987], both 1-D and 3-D models have repeatedly shown that it is not possible for $CO_2$-$H_2O$ atmospheres to produce mean surface temperatures above ~230 K under the relatively faint young Sun [e.g., *Kasting* 1991; *Tian et al.*, 2010; *Forget et al.,* 2013; *Wordsworth et al.,* 2013; *Ramirez et al.,* 2014a]. This is because at above approximately 2 or 3 bars of $CO_2$, the combined effects of intense surface $CO_2$ condensation and Rayleigh scattering outweigh the greenhouse effect and the atmosphere collapses [e.g., *Kasting* 1991; *Forget et al.,* 2013]. Although $CO_2$ ice clouds had originally been thought to warm the climate once cloud cover approaches 100% [e.g., *Forget and Pierrehumbert* 1997], subsequent studies have shown that the warming from this mechanism drastically diminishes at more realistic cloud fractions [*Forget et al.,* 2013]. In addition, the warming becomes nearly null when more accurate radiative transfer methods are employed [e.g., *Kitzmann* 2016]. Furthermore, neither impact-induced cirrus clouds nor runaway greenhouses generate the durations and water amounts necessary to form the valleys [*Ramirez and Kasting* 2017]. It has recently been argued that impact-induced cirrus cloud warming could be more effective if such clouds could have formed in the stratosphere [*Turbet et al.,* 2020]. However, this scenario invokes convection in the $CO_2$ moist adiabat region, which is unphysical because water vapor concentrations are predicted to be very small there [e.g., *Wordsworth et al.* 2013; *Ramirez and Kasting*, 2017]. Furthermore, even if such warming were possible, the warm period would have been far too short to erode the valley networks [*Ramirez and Kasting*, 2017; *Turbet et al.,* 2020]. Although $SO_2$ had been proposed to complement the warming by $CO_2$-$H_2O$ [e.g., *Johnson et al., 2008*; *Mischna et al.,* 2013], this gas is photochemically unstable and rains out of the atmosphere once warm temperatures are approached [e.g., *Tian et al.,* 2010]. Thus, the solution to the early Mars problem had remained rather elusive to this point.

The latest idea to warm early Mars is the $CO_2$-$H_2$ greenhouse solution originally proposed by *Ramirez et al.* [2014a]. Analyses of Martian meteorites suggest that the early Martian mantle was extremely reduced, approaching the iron-wustite buffer, if not lower [e.g., *Grott et al.,* 2011]. Following this idea, *Ramirez el al.* [2014a] had suggested that volcanic outgassing in this scenario would strongly favor reducing gases, including $H_2$, leading to percent level (or higher) concentrations of the gas. In contrast, such a scenario would not work for the early Earth because the mantle had oxidized early on, and thus released $H_2$ concentrations would have been low [e.g., *Trail et al.* 2011]. Nevertheless, this was not the first time that hydrogen had been invoked to warm early Mars. *Sagan and Mullen* [1972] had suggested that Mars could have accreted hydrogen from the protoplanetary disk. However, without a way to replenish the hydrogen, rapid loss to space would occur within just a few million years (Myr) [e.g., *Pierrehumbert and Gaidos* 2011]. In contrast, volcanically-outgassed $H_2$ in the *Ramirez et al*. [2014a] scenario provides a steady source to replenish the escaping hydrogen so long as outgassing rates are high enough, possibly keeping the planet habitable over geologic timescales [e.g., *Ramirez et al.* 2014a]. Plus, high levels of $CO_2$ are possible as reducing gases are oxidized by the products of water vapor photolysis [e.g., *Ramirez et al.* 2014a; *Batalha et al.* 2015].

Furthermore, the *Ramirez et al*. [2014a] study had originally predicted that $CO_2$-$H_2$ collision-induced absorption (CIA) should be even stronger than that for $N_2$-$H_2$. They had argued that preliminary laboratory self-broadening measurements [*Burch et al.* 1969], the stronger greenhouse effect of $CO_2$, along with its great number of modes, all suggested that $CO_2$-$H_2$ absorption should be stronger than that for $N_2$-$H_2$. However, due to the lack of available $CO_2$-$H_2$ CIA cross-sections at the time, these authors had conservatively used $N_2$-$H_2$ CIA as a proxy for $CO_2$-$H_2$ CIA absorption. Motivated by this study, subsequent works, including preliminary estimates [*Wordsworth et al.,* 2017], and laboratory experiments [*Turbet et al.,* 2019], have since verified the prediction of *Ramirez et al*. [2014a], providing more details of the physical mechanisms, and making initial estimates of the $CO_2$-$H_2$ cross-sections. The strength of the revised $CO_2$-$H_2$ CIA means that considerably lower $CO_2$ and $H_2$ pressures were capable of warming early Mars, with $CO_2$ and $H_2$ pressures as low as 0.55 bar and 1% $H_2$, respectively [*Ramirez,* 2017]. Thus, it appears that $CO_2$-$H_2$ may be a very promising solution to the early Mars climate conundrum.

However, whereas using $N_2$-$H_2$ CIA as a proxy for $CO_2$-$H_2$ would underestimate absorption, the preliminary $CO_2$-$H_2$ CIA cross-sections of *Wordsworth et al.* [2017] significantly overestimated the strength of $CO_2$-$H_2$ CIA, as its spectral shape was also different when compared to later laboratory measurements [*Turbet et al.,* 2019]. Nevertheless, the revised CIA of *Turbet et al.* [2019] also have large error bars and thus no reliably accurate $CO_2$-$H_2$ CIA absorption cross-sections exist to date.

Given the uncertainties in $CO_2$-$H_2$ CIA measurements, we will evaluate endmember cases for $CO_2$-$H_2$ CIA absorption to reassess whether a warm early Mars with a $CO_2$-$H_2$ atmosphere is possible using a considerably more complex model, assuming the presence of a northern lowlands ocean. Computed precipitation, runoff and erosion rates will be compared with those inferred from geologic observations to determine two things: 1) Are such warm climates possible? 2) If they are, how well do the predicted runoff and erosion rates agree with current inferred estimates of runoff and erosion for the valley networks and surrounding landscape?

However, even should warm climates be possible, they can certainly be too wet, revealing landscapes that are much more eroded than what are actually observed [*Ramirez and Craddock,* 2018]. Indeed, the valley networks and overall erosion of the Noachian highlands suggest that while the climate was probably relatively warm, the poorly-dissected nature of the valleys indicate a past climate that was drier than many regions on Earth [e.g., *Craddock and Howard,* 2002; *Howard et al.,* 2005]. Thus, our current study attempts to address not only whether warm climates on early Mars are possible, but the nature of such climates as well.

To date, only *von Paris et al.* [2015] have attempted to assess precipitation and runoff rates on early Mars and compared them to valley formation timescales. However, our study differs from that one in several key ways. First, *von Paris et al.* [2015] only assessed $CO_2$-$H_2O$ atmospheres and were unable to obtain warm solutions (as explained above) whereas we assess the $CO_2$-$H_2$ hypothesis of *Ramirez et al*. [2014a] and *Ramirez* [2017], which covers a very different temperature and pressure regime. Secondly, *von Paris et al.* [2015] had used a 1-D radiative-convective climate model whereas our advanced energy balance model (EBM) can also compute latitudinal quantities and include additional aspects like oceans, clouds, the ice-albedo feedback, and seasonal effects. We not only calculate runoff rates, but we also estimate erosion rates for various scenarios. Several new sensitivity studies are also included.

Soto et al. (2010) had initially suggested assessing precipitation rates on early Mars. Although the 3-D study of *Wordsworth et al.* [2015] had assessed rainfall rates on a warm early Mars, they had used grey radiation as a proxy for secondary greenhouse gases. They also do not compute resultant runoff or erosion rates nor compare them with literature estimates. The ice-albedo feedback was not included in their warm simulations. Although they had included a northern lowlands ocean, their model did not assess ocean heat transport and its effects in a dynamic fashion. This is important because ocean heat transport should have been a significant influence on planets located near the outer edge of the habitable zone, like a warm early Mars [*Yang et al.* 2019]. Our topographical assumptions also differ from theirs (as we discuss in the next section).

Another potential candidate for warming early Mars is $CO_2$-$CH_4$ [*Wordsworth et al.* 2017]. However, we do not consider $CO_2$-$CH_4$ collision-induced absorption in this study for a few motives. Not only is $CO_2$-$CH_4$ collision-induced absorption significantly weaker than that for $CO_2$-$H_2$ [*Wordsworth et al*. 2017; *Ramirez and Craddock,* 2018], but it can form anti-greenhouse hazes above a $CH_4$/$CO_2$ ratio of ~0.1 [*Haqq-Misra et al.,* 2008]. $CH_4$ also absorbs strongly at solar wavelengths, which partially offsets its greenhouse effect [*Ramirez et al.,* 2014a; *Ramirez,* 2017]. For these reasons, our study here focuses on $CO_2$-$H_2$ atmospheres, which is (to date) the most potent molecular combination suggested for warming early Mars [*Ramirez et al.,* 2014a; *Ramirez,* 2017].

Therefore, if $CO_2$-$H_2$ fails to warm early Mars and explain valley formation, $CO_2$-$CH_4$ most definitely would.

Our model is the first study to assess the climate on early Mars assuming that the ocean also contributes to the overall heat transport. We will first describe the models and climate modeling procedures for our early Martian $CO_2$-$H_2$ atmospheres in the Methods section. Our final simulations will be summarized in the Results section. We will then discuss the implications of our results in the Discussion section, concluding with final comments.

## 2 Materials and Methods

### 2.1 The Single-Column Radiative-Convective Climate Model

We use a single-column radiative-convective (RC) climate model [e.g., *Ramirez,* 2017; *Ramirez and Kaltenegger*, 2018] to generate the radiative transfer lookup tables that are employed by our energy balance model (see below). The RC model has 55 wavelength bands in the infrared and 38 at solar wavelengths. Our Martian model atmospheres are composed of 100 vertical logarithmically-spaced layers that span from the surface to the ~$5 \times 10^{-5}$ bar pressure level. As water vapor convects in warmer atmospheres, and lapse rates become greater than that for the moist adiabat, lapse rates are adjusted back to the moist adiabatic value. When atmospheres are cold enough to trigger $CO_2$ condensation, the model relaxes to the $CO_2$ adiabat [e.g. *Kasting*, 1991]. Atmospheric expansion as temperatures rise is included in the model [e.g. *Ramirez et al.* 2014a; *Ramirez,* 2017], but this does not have a significant effect for the Martian temperature conditions considered here

The RC model utilizes 8-term HITRAN and HITEMP $CO_2$ and $H_2O$ coefficients, respectively, which are truncated at 500 $cm^{-1}$ and 25 $cm^{-1}$, respectively, computed over 8 temperatures (100, 150, 200, 250, 300, 350, 400, 600 K), and 8 pressures ($10^{-5}$ – 100 bar) [*Kopparapu et al.,* 2013ab; *Ramirez et al.,* 2014ab].

Far wing absorption in the 15-micron band of $CO_2$ employs the 4.3 micron region as a proxy [*Perrin and Hartmann,* 1989]. Analogously, we overlay the BPS water continuum over its region of validity (0 - ~18000 $cm^{-1}$) [*Paynter and Ramaswamy*, 2011]. We also implement $CO_2$-$CO_2$ CIA [*Gruszka and Borysow,* 1997;1998; *Baranov et al.,* 2004; *Wordsworth et al.,* 2010] and $N_2$ foreign-broadening [e.g., *Ramirez et al.,* 2014a; *Ramirez*, 2017]. A standard Thekeakara spectrum is used for the Sun [e.g., *Thekeakara* 1973].

### 2.2 The Mars Energy Balance Model

The main tool for our simulations is the MEBM (Mars Energy Balance Model), which is an updated Mars version of the original EBM used in *Ramirez and Levi* [2018].

The MEBM is similar to other advanced non-grey latitudinally-dependent energy balance models [e.g., *North and Coakley*, 1979; *Willliams and Kasting*, 1997; *Caldeira and Kasting,* 1992; *Batalha et al.,* 2016; *Vladilo et al.*, 2013; 2015, *Forgan*, 2016; *Haqq-Misra et al.,* 2016].  These models are computationally inexpensive and are well-suited for parameter space exploration. Several studies have shown that such advanced EBMs, once properly-calibrated, can produce realistic results comparable to those predicted by 3-D global circulation models (GCMs) [e.g., *Spiegel et al.,* 2009; *Vladilo et al.,* 2015; *Ferreira et al.,* 2014].  These advanced EBMs are of intermediate complexity, including more processes than those that have been considered in single-column radiative convective climate models [e.g., *Ramirez*, 2017; *Wordsworth et al*., 2017], while being faster and easier to use than complex GCMs [e.g., *Forget et al.,* 2013; *Wordsworth et al.,* 2013].

Moreover, we make a number of assumptions which further justify the use of the MEBM for this particular study.  The main one is the assumption of a flat topography for early Mars. Apart from simplicity, this would be approximately true, to first order, for a few reasons.  First, we predict that an ocean occupied much of the northern lowlands, which effectively covers the hemispheric dichotomy seen on the present planet. Secondly, countering the results suggested in *Phillips et al.* [2001], the relatively low cratering density in the Tharsis province suggests that the bulk was not in place prior to valley network formation [*Craddock and Greeley*, 2009]. Recent geophysical models also predict that observed characteristics of purported Noachian ocean shorelines could only be explained if Tharsis was forming during or after valley network formation [*Citron et al.* 2018]. This suggests that the Tharsis region may have been erupting heavily during the late Noachian and early Hesperian period [*Craddock and Greeley*, 2009]. Thirdly, although impact craters can also produce topographical variations, these are both being formed and eroded over time at a very fast rate. Ignoring this factor simplifies our calculations. In effect, the flat topography assumption can be compared with future studies with more complicated topography.

According to the icy highlands hypothesis, large amounts of snow and ice would accumulate on the high altitude regions of Mars, including (and particularly) the Tharsis region [e.g., *Wordsworth et al.,* 2013]. However, this effect would be drastically reduced on a planet without a prominent Tharsis bulge. This is important because the combined effects of the high albedo and thermal inertia of ice and snow can increase the difficulty to simulate warm Martian climates than is otherwise the case [*Forget et al.,* 2013; *Ramirez,* 2017]. A flatter Tharsis region also implies that the rain shadow effect predicted by *Wordsworth et al.* [2015] would not exist – or at the very least – be significantly mitigated under our assumptions, allowing rain to fall on regions like Arabia Terra and Maragitifer Sinus, locations where the geologic evidence supports abundant rainfall in the past [e.g. *Luo,* 2002; *Davis et al.,* 2016; *Davis et al.,* 2019]. We note that recent GCM simulations of the icy highlands hypothesis have been unable to produce results consistent with such geologic observations [*Wordsworth et al.,* 2015]. Thus, we believe that such assumptions not only justify our methods, but are more consistent with observational inferences.

Like other EBMs, the MEBM makes the convenient assumption that planetary heat transfer can be parameterized via diffusion [e.g., *North et al.* 1981; 1983; *Williams and Kasting*, 1997]. As 3-D models have shown, this is not a bad approximation [e.g., *Ferreira et al.* 2014]. The diffusion approximation yields solutions that are consistent with those of more advanced GCMs [e.g., *Spiegel et al.*, 2009; *Ferreira et al.,* 2014].

The MEBM assumes that planets are in equilibrium, emitting as much energy to space as they receive from their host stars (on average). Our model divides Mars into 36 latitudinal bands that are each 5 degrees wide. We express the overall atmospheric-ocean energy balance (thermal, dynamic) by the following equation [*James and North*, 1982; *Nakamura and Tajika*, 2002]:

$$C \frac{\partial T(x,t)}{\partial t} - \frac{\partial}{\partial x} D(1-x^2) \frac{\partial T(x,t)}{\partial x} + OLR - L \frac{\partial M_{col}}{\partial t} = S(1-A) \qquad (1)$$

Where, $x$ is sine(latitude), $S$ is the incoming solar flux, $A$ is the albedo at the top of the atmosphere (within a latitude band), $T$ represents the latitudinally averaged surface temperature, $OLR$ is the outgoing longwave flux, $C$ represents the overall ocean-atmospheric heat capacity, $L$ is the latent heat flux per unit mass of $CO_2$ [$5.9 \times 10^5$ J/kg; *Forget et al.,* 1998], $M_{col}$ is the column mass of atmospheric $CO_2$ that condenses on to the surface or sublimates from the surface to the atmosphere over a given time step, and $D$ represents a calculated diffusion coefficient. We employ a second order finite differencing scheme to solve eqn. 1.

The radiative-convective climate model calculates atmospheric quantities (e.g., temperature, pressure, mixing ratio, and fluxes) that are averaged in the vertical direction while the MEBM computes latitudinal quantities. Whereas some EBMs employ grey radiation, we use the radiative-convective climate model to parameterize relevant quantities, including the stratospheric temperature $T_{strat}(pCO_2, fH_2, T, z)$ outgoing radiation, OLR ($pCO_2$, $fH_2$, T), and planetary albedo A($pCO_2$, $fH_2$, T, z, $a_s$). These variables are a function of the $CO_2$ partial pressure ($pCO_2$), hydrogen mole fraction ($fH_2$), surface albedo ($a_s$), and zenith angle ($z$). Other advanced EBMs have employed complicated polynomial expressions to parameterize such quantities [e.g., *Williams and Kasting,* 1997; *Haqq-Misra et al*., 2016] but these often produce large errors of up to ~20%. The MEBM is instead able to achieve ~99% accuracy through a loglinear interpolation of the relevant quantities ($pCO_2$, OLR, z, A, and $T_{strat}$) over a parameter space spanning $10^{-5}$ bar < $pCO_2$ < 35 bar, 0 < $fH_2$ < 0.2, 150 K < $T$ < 390 K, and 0 < $as$ < 1 for zenith angles between 0 and 90 degrees.

The MEBM can distinguish between different terrains (e.g., land, ocean, and ice) and clouds. The model tracks $CO_2$ cloud formation at each latitude when pressure and temperature conditions allow for $CO_2$ ice to form in the atmosphere. As the atmosphere warms above the freezing point, water clouds appear, which are prescribed an Earth like cloud coverage of 50% [e.g., *Ramirez and Levi*, 2018]. In this latter scenario we assume that rising and subsiding convective motions on a warm planet approximately cancel each other out so that clouds cover approximately half of the planet [*Ramirez*, 2017].

We used Fresnel reflectance data appropriate for ocean reflectivity at different incidence angles [*Kondrat'ev* 1969]. We calculate that ~53% and 47% of the solar energy emission occurs at visible and near-infrared wavelengths. The MEBM then calculates the ice albedo by weighting the average absorption of water ice at visible and near-infrared wavelengths [*Shields et al.*, 2013].

For water ice, we have implemented the following GCM temperature parameterization for the visible and near-infrared albedo of snow/ice mixtures, similar to that in *Curry et al.* [2001] (equation 2):

$$\alpha(visible) = \begin{pmatrix} 0.7 & T \leq 263.15 \\ 0.7 - 0.020(T - 263.15); & 263.15 < T < 273.15 \\ 0.22 & T \geq 273.15 \end{pmatrix} \quad (2)$$

$$\alpha(nir) = \begin{pmatrix} 0.5 & T \leq 263.15 \\ 0.5 - 0.028(T - 263.15); & 263.15 < T < 273.15 \\ 0.22 & T \geq 273.15 \end{pmatrix}$$

Thus, averaged over the near-infrared and visible wavelengths, ice albedo can be as low as ~0.37. Ice albedo is high (~0.6) at temperatures below 263 K, which (crudely) reflects thickening ice cover at colder temperatures. Fresh ice and snow may have even higher albedo than assumed here and very dusty or dirty ice may have somewhat lower values [e.g., North Polar Reservoir Cap; *Bass et al.* 2000]. Our parameterization is a compromise between the two and assumes slightly dirty and mixed snow/ice that thickens at colder temperatures, which is consistent with other similar parameterizations [*Curry et al.* 2001; *Fairen et al.* 2012]. Sensitivity studies (not shown) reveal that assuming a very steep temperature dependence on the visible surface ice albedo for temperatures above 263 K, with a low value (0.35) near the freezing point of water, does little to change the deglaciation thresholds in our Results. This is because the warming is strong enough at these high temperatures to counter the remaining ice cover even if the visible ice albedo is near 0.5.

A typical ice-albedo feedback model is used [*Fairen* et. al., 2012]. We have also created the following exponential fit to the data in *Thompson and Barron* [1981] which relates water ice coverage in an ocean latitude band ($f_{ice}$) to temperature, utilizing a similar approach as that in other advanced EBMs [e.g., *Williams and Kasting* 1997; *Vladilo et al.*, 2013; *Ramirez and Levi,* 2018] (equation 3):

$$f_{ice} = \begin{pmatrix} 1. & ,T \leq 239 \\ 1-\exp((T-273.15)/12.5), & 239 < T < 273.15 \\ 0 & ,T \geq 273.15 \end{pmatrix} \quad (3)$$

We also track the amount of surface $CO_2$ that condenses into ice, melts, or sublimates at a given latitude, which impacts the radiative properties of the atmosphere at each time iteration. We assume that any condensing $CO_2$ accumulates on the surface as dry ice (albedo of 0.6) [*Warren et al.,* 1990], which occurs when the $CO_2$ partial pressure exceeds the $CO_2$ saturation vapor pressure of $CO_2$ ($\Delta pCO_2$) at a given latitudinal band. A similar approach was used in *Haqq-Misra et al.* [2016]. We note that our albedo for dry ice is higher than that used in *Haqq-Misra et al.* 2016 (0.35) although it is similar to that used by *Turbet et al.* (2017). The albedo of $CO_2$ ice replaces that of water ice or land (0.22) in a band once $CO_2$ starts condensing. The $CO_2$ ice thickness $z$ is calculated as follows (equation 4):

$$z = \frac{\Delta pCO_2}{g\rho} \quad (4)$$

Here, the value for acceleration of gravity ($g$) for Mars is 3.73 m/s$^2$ whereas $\rho$ is the density of dry ice (1.6 g/cm$^3$). We also assume that the maximum thickness of latitudinal ice is governed by temperature differences across the ice layer ($\Delta T$) and the geothermal heat flux ($F_g$), $z_{max} = k\Delta T/ F_g$ [*Pollard and Kasting*, 2005]. In the above equation, $k$ is the thermal conductivity of $CO_2$ ice, which has a value of 0.6 Wm$^{-1}$K$^{-1}$ [*Kravchenko and Krupskii*, 1986]. Following *Ramirez et al.* [2014a], we consider that early Mars had a heat flux that is similar to that of the present day Earth or a value of $F_g$ = 0.06 Wm$^{-2}$. Typical $\Delta T$ values are ~ 25 K [e.g., *Pollard and Kasting*, 2005]. These assumptions yield a maximum $CO_2$ ice thickness ($z_{max}$) of 230 m. Once $z_{max}$ in a given latitudinal band is exceeded, glaciers melt along the base and the liquid can flow to lower latitudes before evaporating. However, in agreement with *Haqq-Misra et al.,* [2016], the ice never approaches such thicknesses in our model for these dense $CO_2$ atmospheres either (typically well under 15 m). For comparison, according to equation 4, a 1-bar $CO_2$ atmosphere that completely collapsed would produce an ice layer ~17 m thick. Moreover, some estimates for the geothermal heat fluxes on early Mars are lower that what we assume here, yielding $z_{max}$ values of a few hundred meters or more [e.g., *Fastook and Head, 2015*]. Thus, we consider our $z_{max}$ to be a conservative estimate.

The thermal time scale ($\tau$) of the planetary system (atmosphere, ocean, and land) is calculated at each latitude via equation 5:

$$\tau = \frac{CT}{OLR} \tag{5}$$

Here, $C$ is the zonal heat capacity. Following *Williams and Kasting* [1997], we use the following zonally-averaged effective heat capacity as a function of the ocean and ice fractions ($f_o, f_i$) (equation 6):

$$C = (1-f_o)C_l + fo\{(1-f_i)C_o + f_i C_i\} \tag{6}$$

Following previous work [*North et al.* 1983; *Williams and Kasting* 1997; *Fairen et al.* 2012], we use the following values for land ($C_l$) and ocean ($C_o$) heat capacities: $C_l = 5.25 \times 10^6$, $C_o = 40 C_l$. *Williams and Kasting* (1997) had used one heat capacity value for thin sea ice at temperatures just below the freezing point and another at still lower temperatures. Here, we use a constant $C_i = 2C_l$ for ice.

Cloud albedo is assumed to be a linear function of the zenith angle in radians ($z$) [e.g., *Williams and Kasting* 1997; *Vladilo et al.* 2013] (equation 7):

$$a_c = \alpha + \beta z \tag{7}$$

Where $a_c$ is the cloud albedo, and $\alpha$ and $\beta$ are fitting constants equal to -0.078 and 0.65, respectively. Also, following *Vladilo et al.* [2013], if a negative $a_c$ value is computed, we set $a_c$ to an artificially low value (0.1) although this did not affect our results.

The zonally-averaged surface albedo in our model is determined as follows:

$$a_s = (1-fc)\{(1-f_o)a_l + f_o[f_i a_i + (1-f_i)a_o]\} + f_c a_c \tag{8}$$

Here, $a_s, a_c, a_o, a_i,$ and $a_l$ are the surface, cloud, ocean, ice, and land albedo, respectively. Likewise, $f_c, f_o,$ and $f_i$ are the cloud, ocean, and ice fraction, respectively. Following *Fairen et al.* [2012], at temperatures below 273 K the maximum value between ice and cloud albedo is chosen for $a_s$ to prevent clouds from artificially darkening a bright ice-covered surface.

The MEBM assumes the following heat transfer efficiency parameterization [*Williams and Kasting* 1997]:

$$D = D_o \left(\frac{p}{p_o}\right)\left(\frac{c_p}{c_{p,o}}\right)\left(\frac{m_o}{m}\right)^2\left(\frac{\Omega_o}{\Omega}\right)^2 \tag{9}$$

Earth values are represented by a subscript '$o$'. Here, $p$ is the pressure, $c_p$ is the heat capacity, $m$ is the atmospheric molecular mass, $\Omega$ is the rotation rate, $D$ is the globally-averaged heating efficiency which is recalculated at every time step, and $D_o$ is the globally-averaged heating efficiency for the Earth, which is 0.58 Wm$^{-2}$K$^{-1}$. Thus, for the Earth, $D = D_o = 0.58$ Wm$^{-2}$K$^{-1}$, yielding a latitudinal temperature structure and ice line latitude (~72 degrees) that is similar to those observed for the real planet [e.g.,

*Williams and Kasting* 1997; *Ramirez and Levi* 2018]. For other planets like early Mars, $D_o$ is still 0.58 Wm$^{-2}$K$^{-1}$ but D is scaled in accordance to equation 9 as the aforementioned atmospheric properties vary with time. We perform some sensitivity studies to assess the robustness of our results to moderate changes in $D_o$.

We are specifically interested in testing the hypothesis that early Mars may have had a "warm and semi-arid climate" [e.g., *Ramirez* 2017; *Ramirez and Craddock*, 2018]. Previous studies using similar models have tended to assume fully-saturated atmospheres, which may yield atmospheres that are too moist for this study. For consistency, we chose a sub-saturated Manabe-Wetherald relative humidity profile [*Manabe and Wetherald*, 1967], which is a similar base assumption made by current early Mars GCMs [e.g., *Wordsworth et al.* 2013; *Forget et al.* 2013]. We adopted a Manabe-Wetherald profile over other common sub-saturated profiles ( tropospheric RH = 50%) because it treats variations of RH with height more realistically, impacting tropospheric distributions of water vapor and precipitation (Figure 1). Nevertheless, a Manabe-Wetherald profile does not yield answers that are significantly different on average from those of a RH = 50% profile [*von Paris et al.* 2015].

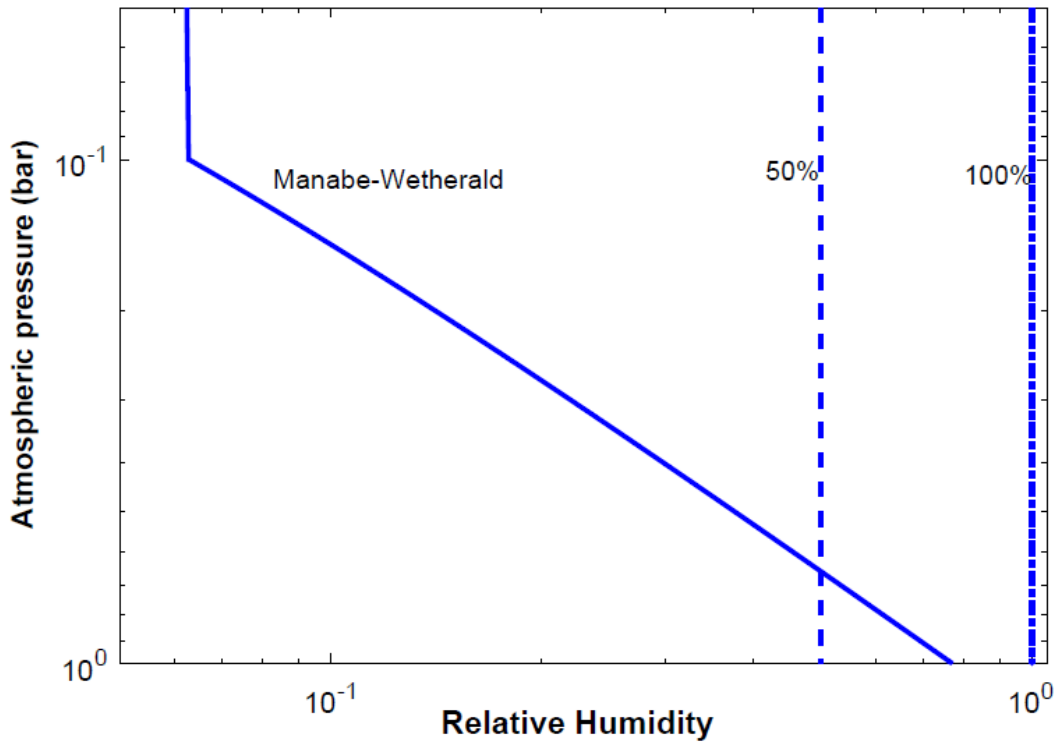

**Figure 1:** Manabe-Wetherald (solid blue), 50% (dashed), and fully-saturated (dashed-dotted) relative humidity profiles for comparison. The Manabe-Wetherald profile relative humidity linearly decreases from ~77% at the surface to the top of the convective layer above which the relative humidity is assumed to be constant.

The MEBM adopts an explicit forward marching numerical scheme that self-consistently computes model parameters over the year. A constant time step is used (i.e., some fraction of a day). We can vary planetary and stellar inputs, including star type, stellar flux, stellar class, planetary tilt, orbital eccentricity, and planetary rotation rate. The MEBM can compute annual-averaged computations [e.g., *Ward et al.* 1974] although we calculate the entire seasonal cycle in this study. The model integrates over time and latitude and convergence is reached when the variation in mean annual planetary surface temperature becomes less than some value (0.02 K for this study).

*2.3 Precipitation and ocean parameterizations and their dependence on ocean area*

Precipitation is difficult to simulate for climate models, especially for planets different from the Earth. Some models employ precipitation thresholds for the atmospheric water content, above which precipitation occurs [e.g., *Urata and Toon*, 2013; *Wordsworth et al.* 2013]. Others assume that clouds simply rain out all of their water [e.g., *Segura et al.* 2008] or employ precipitation efficiencies derived from Earth and applied to Mars [*von Paris et al.,* 2015]. We instead, parameterize water precipitation (P) via the following [*Pollard and Kasting*, 2005](equation 10):

$$P = fac \frac{\rho_a h_q q_a}{\tau_p} r_a^3 \qquad (10)$$

Here, $\rho_a$ is the atmospheric density, $h_q$ is the atmospheric water vapor thickness, equivalent to the highest altitude from which precipitation (rain or snow) can still reach the surface, $q_a$ is the surface specific humidity, $r_a$ is the surface relative humidity (0.77 for Manabe-Wetherald), *fac* is an ocean size scaling factor, and $\tau_p$ is a precipitation timescale. This equation is derived by integrating the column mass of water vapor in the atmosphere, assuming that the water is removed by the degree of saturation [*Pollard and Kasting*, 2005]. The model assumes precipitation falls as rain for latitudinal temperatures above 273 K, falling as snow otherwise. This parameterization has been shown to obtain a relatively close match of the latitudinal distribution of precipitation on the Earth [*Pollard and Kasting*, 2005]. On Earth, ~ 94% of the trospospheric water vapor is concentrated within the first 5 km [e.g., *Tomasi,* 1994]. Thus, we assume that $h_q$ is ~ 6 km for Earth, the same value used by *Pollard and Kasting* [2005], which is also supported by radar measurements [e.g*., Short et al.* 1997; *Geerts and Dejene* 2005]. This value of $h_q$ is approximately 3 times Earth's water vapor scale height [*Weaver and Ramanathan*, 1995], which captures more than 97% of Earth's tropospheric water vapor. For early Mars, we also assume that $h_q$ is 3 times the computed water vapor scale height over a given latitude band. For example, our model predicts a mean water vapor scale height of ~2.8 – 2.9 km for typical warm Martian atmospheres (~280 K mean surface temperature), yielding an $h_q$ of ~ 8.4 – 8.7 km, although values considerably higher or lower than this are possible at any given latitude band. We calibrate $\tau_p$ in our Earth model so that equation 10 yields ~2.65 mm/day [e.g. *Xie and Arkin*, 1997], which is the average terrestrial precipitation rate. In the MEBM, this corresponds to $\tau_p$ = 13.5 days, which is similar to the value assumed in *Pollard and Kasting* (2005) and only slightly larger than what has been inferred from observations (~ 9 - 10 days) [e.g., *van der Ent and Tuinenberg* 2017]. That said, we do not know what $\tau_p$ would be for a warm early Mars as this largely depends on unknowable factors like the exact ocean/continental configuration, topography, atmospheric-ocean circulations and atmospheric composition. However, given that $\tau_p$ on Earth is only 2 days shorter over oceans than on land [*van der Ent and Tuinenberg* 2017], we have reason to believe that $\tau_p$ varies no more than ~20 - 25% across semi-arid to wet climates, with drier ones likely having slightly larger precipitation timescales. For these reasons, in addition to simplicity, we keep $\tau_p$ = 13.5 for all calculations.

We also perform sensitivity studies of ocean size and how that influences the resulting precipitation, runoff, and erosion rates to selected valley network regions. We note that equation 10 implicitly assumes an Earth-like planet with 70% ocean coverage. However, average precipitation amounts across the planet should be lower for planets with relatively small ocean fractions. In our model *fac* scales precipitation rates by the ocean area divided by 70% of the surface area of Mars. Our baseline ocean is that by *Di Achille and Hynek* [2010], which is ~ 500 m global equivalent and covers nearly 36% of the planetary surface area (i.e., *fac* = 36/70). We also perform limited comparisons against two smaller oceans (discussed in Results). Thus, we effectively assume that the ocean is able to maintain a constant relative humidity profile around the planet although (in reality) this should vary with location. Besides this being a common assumption in models [e.g., *Ramirez et al.* 2014a; *von Paris et al.* 2015; *Wordsworth et al.* 2017], the warm atmospheres we consider here are very dense, heat transfer is efficient, and the topography is flat. Even though the Earth has significant topographical variations, and the atmosphere is less dense than those of our warm simulations (see Results), satellite observations suggest that seasonal variations in tropospheric relative humidity only vary from ~0 – 10% between the northern and southern hemisphere [*Peixoto and Oort*, 1996]. So, for all of these reasons we suspect (but cannot prove) that mean relative humidity spatial variations would only have a second order effect on our simulations (although we elaborate further in the Discussion section).

Furthermore, we note that ocean size (particularly area, assuming an ocean that is deep enough) and relative humidity are directly correlated. This is because evaporation rates (in mass/time) are larger for water bodies of larger surface area. Thus, average relative humidity and surface temperature on planets with smaller oceans should both be lower. However, nearly all of the warming in these dense $CO_2$-$H_2$ atmospheres is from $CO_2$-$H_2$, not water vapor [*Ramirez et al.* 2014a], with mean surface temperatures being only a couple degrees higher in fully saturated atmospheres versus much drier (RH =0 and 50%) ones [*Ramirez et al*. 2014a; *Ramirez*, 2017]. Thus, to first order, this justifies the use of the same relative humidity profile for this analysis. As we have said, however, ocean area has a large effect on precipitation rates, which we do take into account and show later in Results.

Moreover, it is possible that parts of the ocean are too cold to contribute to the hydrologic cycle at any time during the year. We assume here that water vapor does not evaporate from latitudes where the ocean temperature is below the freezing point of water. The assumption here is that evaporation rates below 273K are effectively zero, which, to first order, is true [e.g., *Linacre,* 1977]. When that happens, we scale global precipitation rates down by the fractional area of the ocean that exhibits sub-freezing temperatures. This ensures that 1) global precipitation rates are zero when the mean surface temperature of the ocean is below zero and 2) the global precipitation is maximum when the entire ocean is warm. Nevertheless, this does not necessarily mean that the ocean is frozen when temperatures are below 273 K. Thin ice and open regions of ocean can still exist at temperatures that are several degrees below the freezing point of water [e.g., *von Deimling et al.,* 2006; *Annan and Hargreaves*, 2013*; Ramirez and Levi,* 2018].

### *2.4 Runoff rate parameterizations*

Where there is land, we also estimate runoff rates as a function of latitude. Runoff rates are determined from precipitation minus evaporative and transmission losses into the regolith (i.e. infiltration). On Earth, transmission losses are determined by quantifying the water balance via stream gauges at different locations within a reach [e.g., *Schoene*r, 2016]. Complex regression equations and models can be constructed from such data [e.g., *Cataldo et al.* 2010]. For the unknown conditions of early Mars, we employ equation 11:

$$R_{run} = P - T - E \qquad (11)$$

Where $R_{run}$ is the runoff rate, $T$ is the transmission loss into the regolith, and $E$, evaporative losses. Transmission losses for different terrains vary greatly, ranging from ~20% to up to 95% of the total precipitation, P [e.g., *Greenbaum et al.* 2002; *Lange*, 2005; *Goodrich et al.,* 2004]. However, an accepted average transmission loss over a long period of time is approximately 30– 40% [*Schoener,* 2016]. Here, we assume a 35% mean annual transmission loss ($F$= 0.35) for all cases. The evaporation rate can be written as the following [e.g., *Ramirez et al.* 2014b](equation 12):

$$E = C_d u \rho_s q (1 - RH) \qquad (12)$$

Here, $C_d$ is the surface drag coefficient, $u$ is the near-surface wind speed, $\rho_s$ is the atmospheric density, and $RH$ is the near-surface relative humidity. We assume standard values for $C_d$ of $1.5 \times 10^{-3}$ [e.g., *Hidy*, 1972; *Pond et al.* 1974] and $u$ of 5 m/s [e.g., *Pond et al.* 1973], respectively. The latter assumption for $u$ assumes that thick and warm early Mars atmospheres have similar wind speeds as the Earth. The MEBM computes the remaining quantities. Thus, equation 11 can be written as (equation 13):

$$R_{run} = P(1-F) - E \qquad (13)$$

$R_{run}$ represents the model computed discharge rates (in mm/day). The MEBM also allows seasonal snowmelt to be included in $R_{run}$. We compare $R_{run}$ versus those inferred in the literature from geologic observations. When known, discharge rates ($Q$) in streams can be determined by the product of flow velocity ($V$) and the channel surface area ($A$) in Darcy's equation. This can be expanded to include a few more variables in the popular Manning's equation variant (equation 14).

$$Q = VA = VHW = H^{5/3} S^{1/2} g^{1/2} W n^{-1} \qquad (14)$$

Here, $H$ and $W$ are the channel height and width, respectively, $S$ is the slope, $g$ is gravity, and $n$ is the Manning roughness coefficient. However, in the case of the Martian valley networks, the values for $S$ and $n$ will be unknown. In place of this, empirical functions that compute $Q$ by relating it to $W$ were derived by averaging the characteristics of streambeds in the Missouri River basin [*Osterkamp and Hedman*, 1982], and this approach has been applied previously to estimating discharges in channels identified in Martian valley networks [*Irwin et al.*, 2005]. Defining $Q$ is very difficult, as it depends on the geologic properties of the streambed, including rock/soil type and grain size. The exact expression differs on the frequency and intensity of any flood events that may occur within a region [*Osterkamp and Hedman*, 1982]. Assuming that flood events control alluvial channel dimensions in humid regions, some authors have used such expressions to compute the runoff rate for Martian valleys [e.g., *Irwin et al.* 2005]. However, not only are the recurrence intervals of any flood events on early Mars unknown [*Irwin et al.* 2005], recurring floods that are too large and frequent may overestimate overall discharge in semi-arid climates [*Barnhart et al.* 2009]. Thus, we instead start with the following expression for the mean discharge rate, which includes the seasonal warming that the MEBM predicts and is a more appropriate $Q$ to compare against MEBM results [*Osterkamp and Hedman*, 1982] (equation 15):

$$Q = CW^{1.71} \qquad (15)$$

Here, $C = 0.027$. However, $Q$ needs to be scaled for the lower gravity of Mars, which results in larger channel widths for the same $Q$ [e.g., *Irwin et al.* 2005], requiring $C$ to decrease in response without changing the exponent. *Moore et al.* (2003) proposed that gravity is scaled to the -0.23 power in expressions of Q. Thus, equation 15 is multiplied by $((3.73/9.81)^{-0.23})^{-1.71}$, which equals ~ 0.68, yielding the final expression we use for the discharge in Martian valley streams (equation 16):

$$Q = 0.018W^{1.71} \tag{16}$$

This last expression is the same one used in *von Paris et al.* (2015). Finally, dividing *Q* by the catchment or drainage area yields runoff rates that can be compared with $R_{run}$ derived from the MEBM. Given the uncertainties in these expressions, the numbers here may get within a factor of 2 or 3 of the actual runoff rates [e.g., *Howard et al.* 2005].

*2.5 Erosion rate parameterization*

Given the MEBM computed runoff rates, we also attempt to estimate erosion rates and compare those values to estimates provided in the literature. We employ the universal soil loss equation (USLE)[*Wischmeier and Smith,* 1978](equation 17) minus agricultural terms (e.g., organic content) to estimate topsoil erosion rates on Mars:

$$A = FRKLS \tag{17}$$

Where *A* is the average annual soil loss (in kg/m$^2$), *F* is the conversion factor from tons/acre to kg/m$^2$ (0.22417), *R* is the rainfall erosivity index, *K* is the soil erodibility factor, and the product *LS* is the topographic factor (which depends on slope and basin slope length). The rainfall erosivity (*R*) index is determined by an annual integration of the product of daily rainfall kinetic energy and precipitation rate. However, storm kinetic energy on early Mars cannot be predicted by the MEBM, as it requires specific knowledge of the individual storm's characteristics, including the size distribution, frequency, and velocity of raindrops [e.g., *Petru and Kalibova* 2018]. Instead, we assume that storm kinetic energy on early Mars is the same as that for the Earth and that *R* is only a function of the annual precipitation rate. Assessing erosion rates of impacts on early Mars, *Segura et al.* [2008] had utilized equation 18 for *R* [*Lo*, 1985]:

$$R = \frac{3.48X + 38.46}{1.702} = 2.04X + 22.596 \tag{18}$$

Here, *X* is the annual precipitation rate (in cm/year) and *R* has been converted such that *A* yields the correct units [*Segura et al.,* 2008]. For consistency, we will use the same equation. In reality, *R* should vary throughout the planet following regional variations in climate and soil properties [e.g., *Renard et al.,* 1994; *Pham et al.,* 2001]. As we show in Results, however, the climate may have been relatively homogeneous in these dense early Martian atmospheres, which itself is consistent with the general distribution of modified impact craters [e.g., *Craddock and Howard*, 2002]. The topographic factor (*LS*) varies between ~0 for zero degree slopes to ~13 for a 20 degree slope that is 1000 ft long [*Wischmeier and Smith*, 1978**].** Previous work had assumed *LS* =5, which may be a reasonable value for large basins in the Chattahoochee river basin (~>50 mi$^2$) [*Faye et al.*, 1979; *Segura et al.,* 2008], corresponding to a slope of 5- 10 degrees or even more [*Wischmeier and Smith*, 1978]. However, valley network slopes over great portions of their lengths tend to be ~1 degree or less [*Hoke et al.* 2011; *Orofino et al.* 2018]. Here, we assume that *LS =0.5,* which is consistent with a significantly smaller slope [*Wischmeier and Smith*, 1978]. The soil erodibility factor (*K*) is a strong function of the chemical and physical processes that had operated on early Mars. The dominant erosional processes would have been impacts and chemical weathering because there is little evidence of physical weathering processes, like freeze-thaw cycles, on early Martian terrains [e.g., *Grotzinger et al.* 2015; *Davies et al.* 2016]. In the absence of plate tectonics, this would suggest that the regolith could have been relatively soft, consisting of easily erodible materials (e.g. alluvium, sand, and silt). In that case, *K* would have been rather high. For this end member case, we assume a *K* of 0.5. However, such a high *K* neglects volcanism, which, on an active early Mars, would be constantly supplying fresh volcanic materials that solidify into rock and partially counter erosion. In this latter case, we assume that the regolith is a mix of soil and rock fragments. A *K* value of 0.3 is representative of a somewhat typical *K* for terrestrial soils [*Wischmeier and Smith*, 1978; *Faye et al.,* 1979]. If we assume a 50% rock by volume fraction, *K* halves to 0.15 [e.g., *Rennard et al.* 1997]. We use *K* = 0.15 as our lower bound. These *K* values assume no organic matter present, which is reasonable to suggest for early Mars. Nevertheless, erosion rates are notoriously difficult to predict [e.g., *Craddock and Howard,* 2002; *Golombek and Bridges* 2000; *Segura et al.* 2008] and thus, our computed erosion rates yield (at best) order of magnitude certainty. However, we have estimated these values in an effort to validate the model output with what is known about the surface geology on Mars.

## *2.6 Climate modeling procedures*

The MEBM is used to test whether a warm and semi-arid climate capable of producing enough rain to form the valley networks and explain the observed fluvial erosion can be achieved with the $CO_2$-$H_2$ atmospheres initially proposed by *Ramirez et al.* [2014a] and expanded in *Ramirez* [2017]. We assume an average solar flux value of 439 W/m$^2$, appropriate for conditions ~ 3.8 Ga at a semi-major axis distance of 1.52 AU. We also assume present values of obliquity (25 degrees), eccentricity (0.0934), and land albedo (0.22). However, if surface albedo was lower in the distant past when erosion rates were low and soil was not as available [*Mischna et al.,* 2013], somewhat higher surface temperatures may be possible. Nevertheless, erosion rates were high during valley network formation, so we assume that soil then was similar to today's.

As per our earlier works [*Ramirez et al.* 2014a; *Ramirez*, 2017], modeled atmospheres are assumed to be mainly $CO_2$ (90 – 99%) with a smaller amount of $H_2$ (1 – 10%). As discussed before [*Ramirez et al.*, 2014a], we assume that volcanic outgassing rates on early Mars were comparable to those on present day Earth, on a per unit basis [*Montesi and Zuber* 2003]. We also argue for a reduced early Martian mantle, which would have favored the outgassing of reduced gases [*Grott et al.* 2011]. Hydrogen concentrations of 10% and above may be achievable because both magnetic fields and the assumption of a fully spherical planet (over a 1-D representation) may help slow down hydrogen escape [*Stone and Proga*, 2009; *Ramirez et al.* 2014a]. These concentrations also assume that hydrogen escapes at the diffusion limit, which is the highest escape rate for these conditions [e.g., *Hunten* 1973]. Any leftover atmospheric gas is $N_2$.

As mentioned earlier, given the large errors bars on recently-estimated $CO_2$-$H_2$ CIA [e.g., *Wordsworth et al.*, 2017; *Turbet et al.,* 2019], there currently remains no reliable absorption cross-sections for $CO_2$-$H_2$. In this study, we perform two sets of calculations. As we had done before [*Ramirez et al.,* 2014a], a lower bound absorption estimate can be obtained by assuming that $CO_2$-$H_2$ CIA is of the same strength as $N_2$-$H_2$. Likewise, an upper bound on $CO_2$-$H_2$ CIA is possible via the *Wordsworth et al.* [2017] cross-sections. We discuss the implications of using both estimates.

## 3 Results

### *3.1 Typical Martian temperature distributions and comparison to Earth*

The atmospheric and ocean heat transfer on a warm early Mars with a large northern lowlands ocean vary significantly across the planet. Seasonal temperatures are drastically reduced over the ocean (a few degrees) whereas minimum and maximum temperatures deviate from the mean by up to 40 – 50 degrees near the south pole (Figure 2). The large differences in seasonal temperatures are attributable to the relatively low heat capacity of land and ice with respect to the ocean [e.g., *Spiegel et al.* 2009]. In spite of the higher maximum and lower peak temperatures near the south pole, however, mean annual surface temperatures are still highest near and around the equator close to the ocean shoreline (Figure 2). Whereas large variations in seasonal temperatures exist over land, the equator-pole mean annual temperature gradient is relatively small (< 20 K) in both cases. In comparison, the equator-pole mean annual temperature gradient on our model Earth is ~45 K, with seasonal variations no higher than ~10 – 12 K (Figure 3).

According to equation 1, equator-pole temperature gradients are impacted by the diffusion parameter (*D*), which is proportional to atmospheric pressure and heat capacity. Early Mars, which is farther away from the Sun than our planet is, needs a thicker atmosphere and stronger greenhouse effect to achieve a similarly warm mean surface temperature as the Earth. In other words, whereas the magnitude of the greenhouse effect (above the blackbody emitting temperature) on Earth is ~33K, at least ~60K or 70K would be needed to warm Mars. These atmospheres are also slightly cooler than the Earth (272 K and 280 K vs. Earth's average mean surface temperature of 288 K) and so less atmosphere is needed to reduce latitudinal temperature contrast relative to an even warmer planet. Thus, net meridional heat fluxes should be significantly smaller on a warm early Mars than the Earth (Figure 4).

Mean temperatures are low enough at higher southern latitudes that an ice cap forms for part of the year in the 1 bar $CO_2$ 10% $H_2$ case (ice line extent: 60°S). Also, only part of the ocean exhibits slightly sub-freezing temperatures in the 10% H$_2$ scenario, whereas the rest remains warm enough to contribute to the hydrologic cycle (Figure 2a).

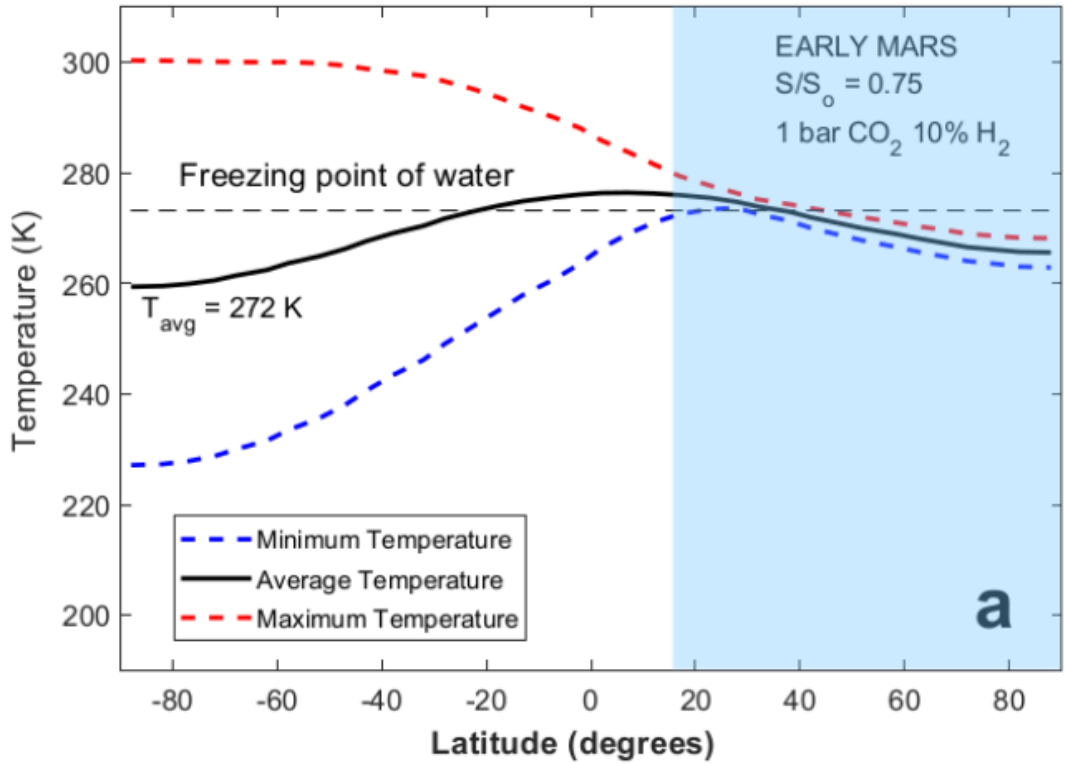

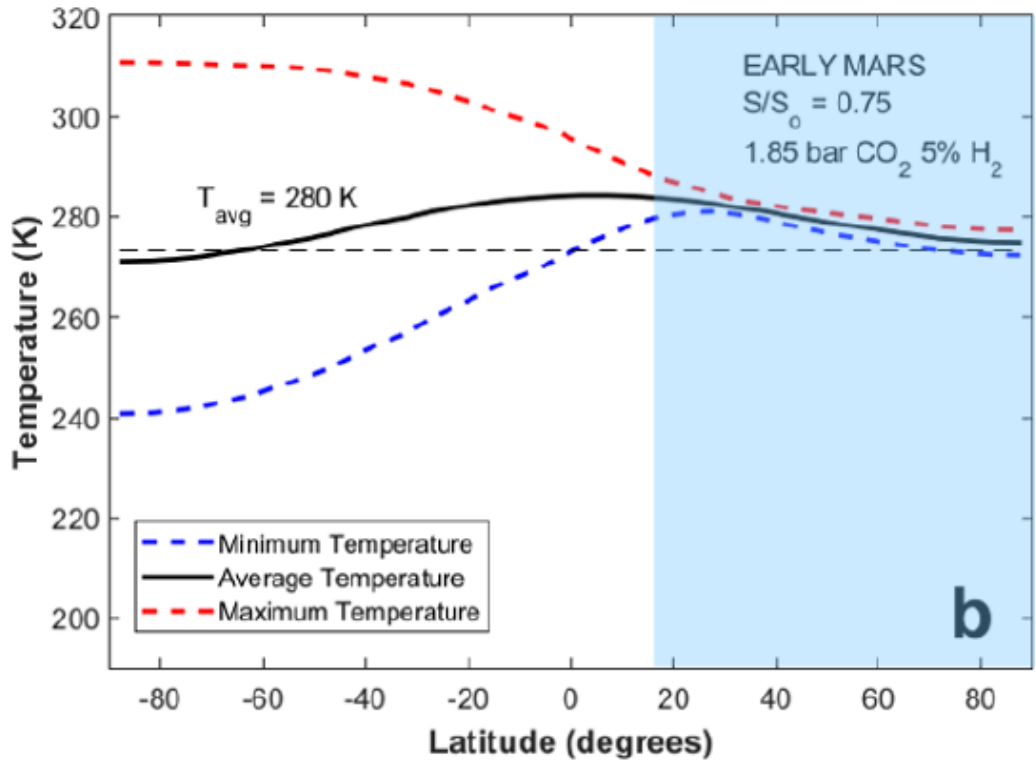

.

**Figure 2:** Latitudinal mean annual temperature distributions for (a) 1 bar $CO_2$ 10% $H_2$ and (b) 1.85 bar $CO_2$ 5% $H_2$ atmospheres. Average annual temperatures(black) and minimum (blue dashed) and maximum (red dashed) seasonal temperatures are also shown. The baseline ocean of *DiAchille and Hynek* [2010] (light blue shaded region) was assumed in these calculations

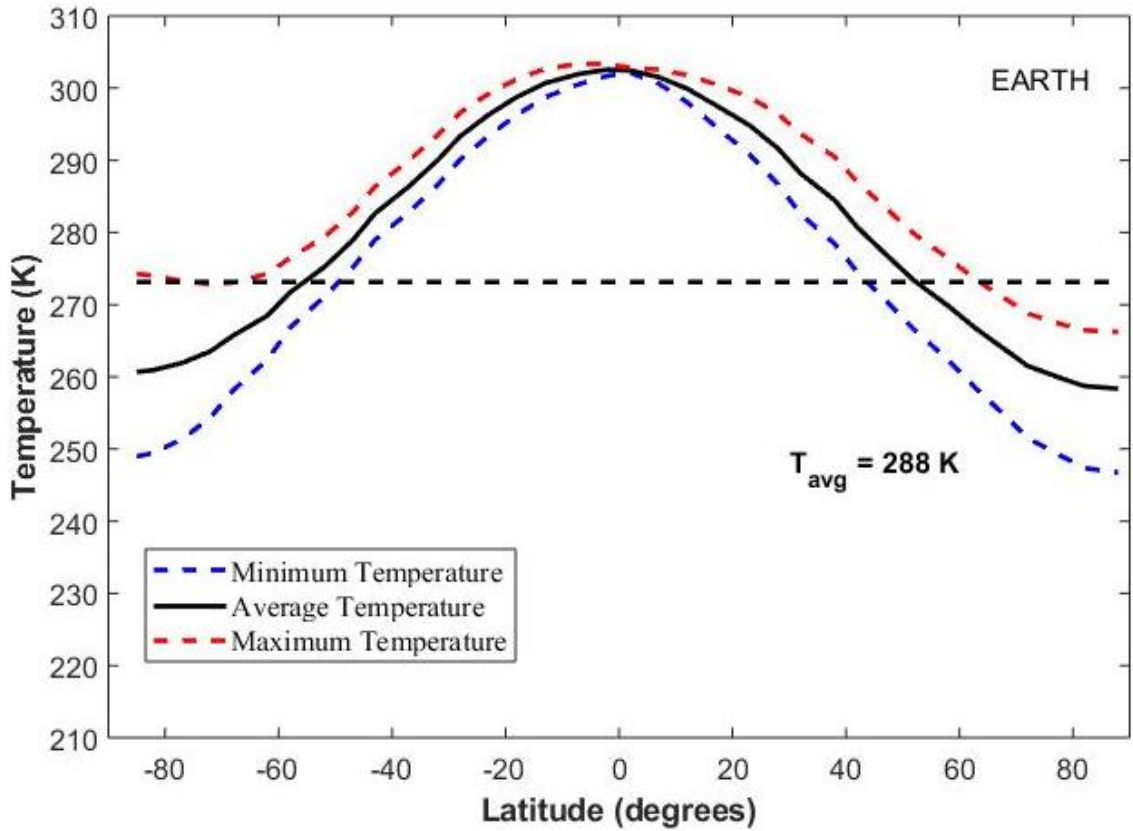

**Figure 3:** Latitudinal mean annual temperature distributions for the Earth. Average annual temperatures (black) and minimum (blue dashed) and maximum (red dashed) seasonal temperatures are also shown. The atmosphere for our model Earth consisted of 1 bar $N_2$ with 330 ppm $CO_2$, following previous work [e.g. *Kasting et al.* 1993]. Ocean and continental fractions were determined using the data of *Kossinna* [1921]. A flat topography was assumed.

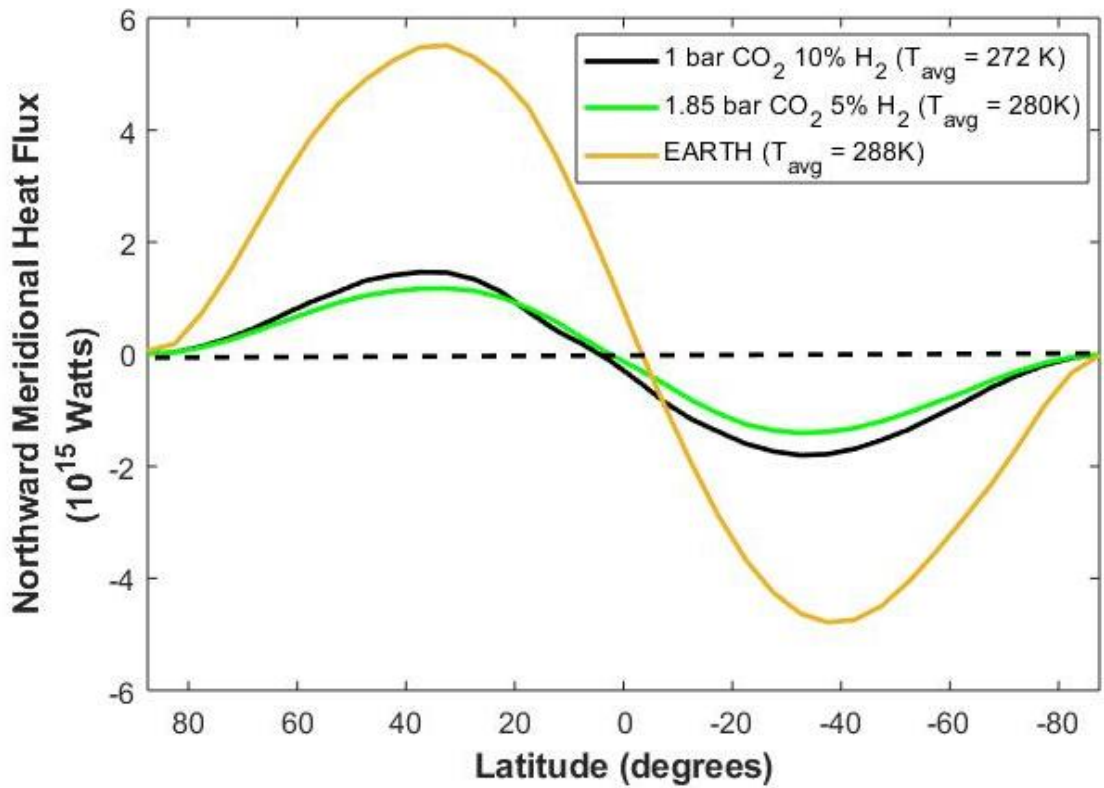

**Figure 4:** Total annual northward atmospheric and oceanic heat fluxes for the Earth (brown) and 2 dense $CO_2$-$H_2$ early Martian atmospheres for comparison. Both the 1 bar $CO_2$ 10% $H_2$ (black) and 1.85 bar $CO_2$ 5% $H_2$ Martian atmospheres have smaller meridional heat fluxes than the Earth. The baseline ocean of *DiAchille and Hynek* [2010] was assumed for the Mars calculations whereas the real world continental and ocean fractions at each latitude band were employed for Earth. The Earth case agrees very well with the similar calculation made by the *Williams and Pollard* [2003] GCM.

*3.2 Warm vs. Cold start study*

Some authors argue that early Mars was more easily warmed starting from a perpetually cold and icy climate warmed by infrequent warming events dominated by any of a number of mechanisms, including snowmelt [e.g., *Wordsworth et al.,* 2013; *Wordsworth et al.* 2015*; Forget al.,* 2013], limit cycles [*Batalha et al.* 2016], and meteoritic impacts [e.g., *Urata and Toon,* 2013]. In contrast, others believe that valley network formation was more easily achieved with a warmer baseline climate capable of producing rain [e.g., *Craddock and Howard*, 2002; *Ramirez et al.* 2014a*; Ramirez*; 2017; *Luo et al.* 2017; *Ramirez and Craddock*, 2018]. We perform cold and warm start simulations to evaluate both possibilities. We would also like to gauge how icy early Mars can be and still be a warm planet. This was performed first with $CO_2$-$H_2$ CIA (Figure 5) and then later with $N_2$-$H_2$ CIA (Figure 6) For both cold and warm starts, $CO_2$ and $H_2$ greenhouse gas pressures were gradually increased until either mean surface temperatures exceeded the freezing point of water or the atmosphere had collapsed.

For the cold starts, we assumed a starting surface temperature of 210 K and a frozen northern lowlands ocean. Temperatures may rise and while temperatures are cold enough to form ice, it forms on part of the latitude band. In comparison, the starting surface temperature for warm starts was assumed to be 280 K. The baseline land ice fraction for both cold and warm starts is assumed to be 50%, unless stated otherwise, if the land surface is cold enough to form ice (although we assess higher and lower land ice fractions as well). Latitude bands that are warm enough are completely ice-free. With the strong $CO_2$-$H_2$ CIA assumption [*Wordsworth et al.* 2017], warm starts yield warm solutions. For an atmosphere with 2.9 bars of $CO_2$, only 1% $H_2$ is needed to achieve warm mean surface temperatures (Figure 5a). Likewise, similar results were found for the 1bar $CO_2$ 10% $H_2$, 1.85 bar $CO_2$ 5% $H_2$, and 1.9 bar $CO_2$ 3% $H_2$ cases. The 3%, 5% and 10% $H_2$ cases all require under 2 bars of $CO_2$, which satisfies recently suggested paleopressure constraints [e.g., *Kite et al.* 2014; *Hu et al.* 2015; *Craddock and Lorenz* 2017; *Kurokawa et al.,* 2018] (Figure 5a). These numbers also corroborate recent results from 1-D radiative-convective climate modeling *[Ramirez,* 2017; *Wordsworth et al*., 2017]. Nevertheless, because of the ice-albedo feedback, temperatures are in general a couple to a few degrees cooler than for the equivalent case from the original 1-D radiative-convective climate calculations [*Ramirez*, 2017].

In contrast, the atmosphere for the 2.85 bar $CO_2$ 5% $N_2$ case collapsed with a mean surface temperature of ~208 K (Figure 5a). The surface remained frozen even with further increases in surface pressure. For comparison, the mean surface temperature for this case from the original radiative-convective modeling calculation was ~230 K [*Ramirez et al.,* 2014; *Ramirez*, 2017]. Thus, the MEBM ice-albedo feedback reduces predicted mean surface temperatures for $CO_2$-$H_2O$ atmospheres, especially with a negligible greenhouse effect from $CO_2$ ice clouds.

The numbers with cold starts are somewhat different, although $CO_2$-$H_2$ absorption helps deglaciate an initially cold planet. A warm solution can be achieved in the 2.9 bar $CO_2$ 1% $H_2$ case if the land ice fraction does not exceed ~50% (Figure 5b). Thus, at land ice fractions less than 0.5, deglaciation is possible. This abrupt transition to higher temperatures is due to the sudden decrease in surface albedo following a massive greenhouse effect that deglaciates an icy planet [*Yang et al.,* 2017]. At even higher $H_2$ concentrations, planets can deglaciate from an initially ice state. Perhaps surprisingly, even in the 1.85 bar $CO_2$ 5% $H_2$ case, the greenhouse effect was strong enough to deglaciate a completely ice-covered planet (Figure 5b). Thus, both cold and warm start scenarios can achieve similarly warm surface temperatures with the same amount of greenhouse gases if land ice fractions are not too high (Figure 5).

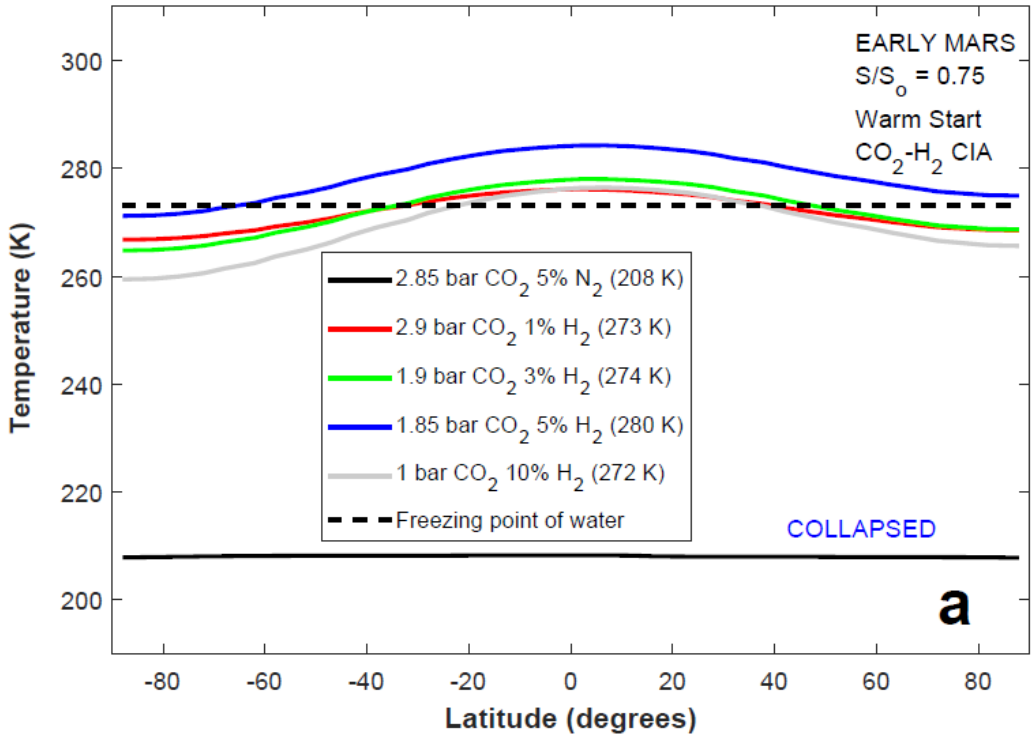

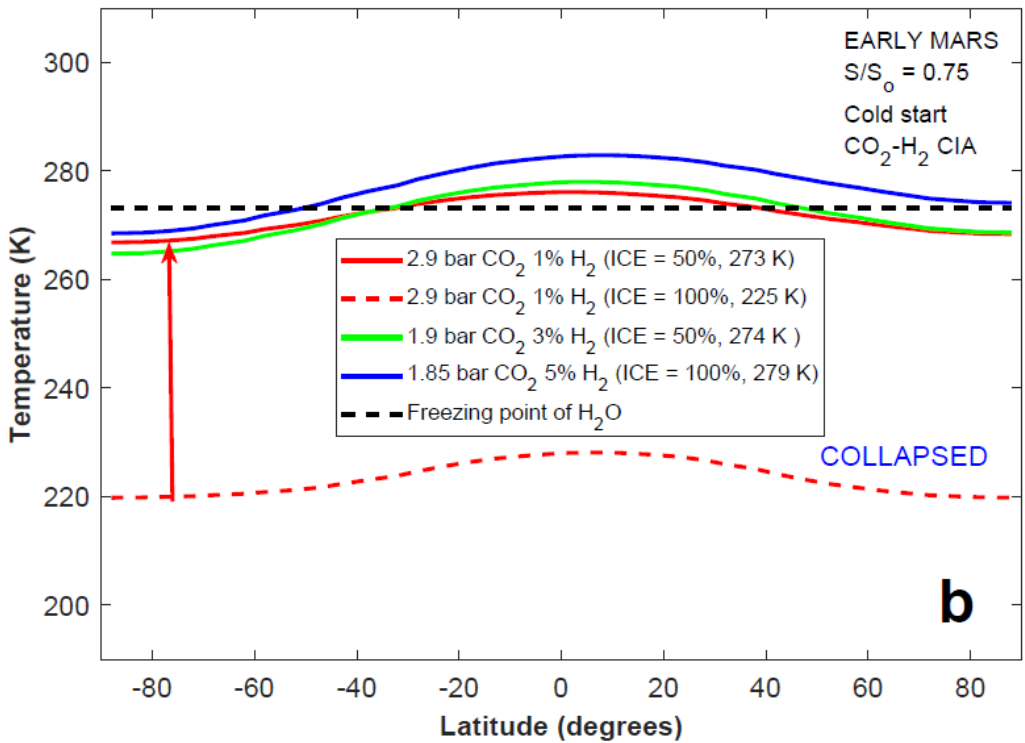

**Figure 5:** Latitudinal mean annual temperature distributions for (a) warm and (b) cold start cases with $CO_2$-$H_2$ CIA. The 5% $N_2$ (black), 1% (red), 3% (green), 5% (blue) and 10% (grey) $H_2$ cases are shown for various $CO_2$ levels and different land ice fractions. Collapsed atmospheres in Figure 5b are shown as dashed lines.

In contrast, the ice-albedo feedback is more effective with the $N_2$-$H_2$ CIA proxy assumption [*Ramirez et al.* 2014a] (Figure 6). For warm starts with $H_2$ concentrations of < ~5% $H_2$, warm solutions are not possible. This also agrees with previous radiative-convective climate modeling simulations [*Ramirez et al.* 2014a], although as mentioned earlier, slightly higher greenhouse gas pressures are needed in the MEBM to achieve the same surface temperatures. The situation is slightly different for cold starts. A warm solution is not possible for the 2.95 bar $CO_2$ 5% $H_2$ cold start case for land ice fractions above 0.51 (Figure 6b), although deglaciation is possible at land ice fractions of 0.5 and below. However, for the 10% $H_2$ (2.83 bar) cold start case, warm solutions are achievable even on a completely glaciated planet (Figure 6b). In total, these latter CIA results support the notion that the albedo of surface ice presents a significant challenge in warming an initially cold and icy planet for atmospheres with $H_2$ concentrations as high as ~5% [*Ramirez*, 2017; *Ramirez and Craddock*, 2018]. Although we find that the ice problem can be circumvented at significantly higher $CO_2$ and $H_2$ concentrations by overestimating $CO_2$-$H_2$ CIA, the ice-albedo feedback becomes an issue under more modest CIA assumptions.

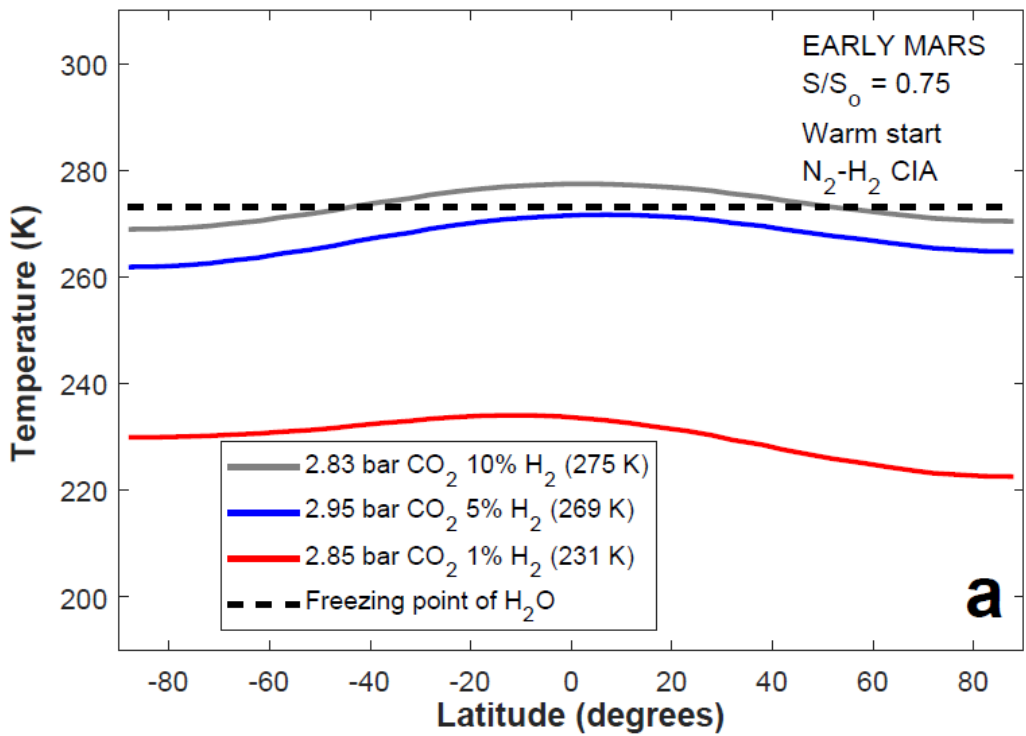

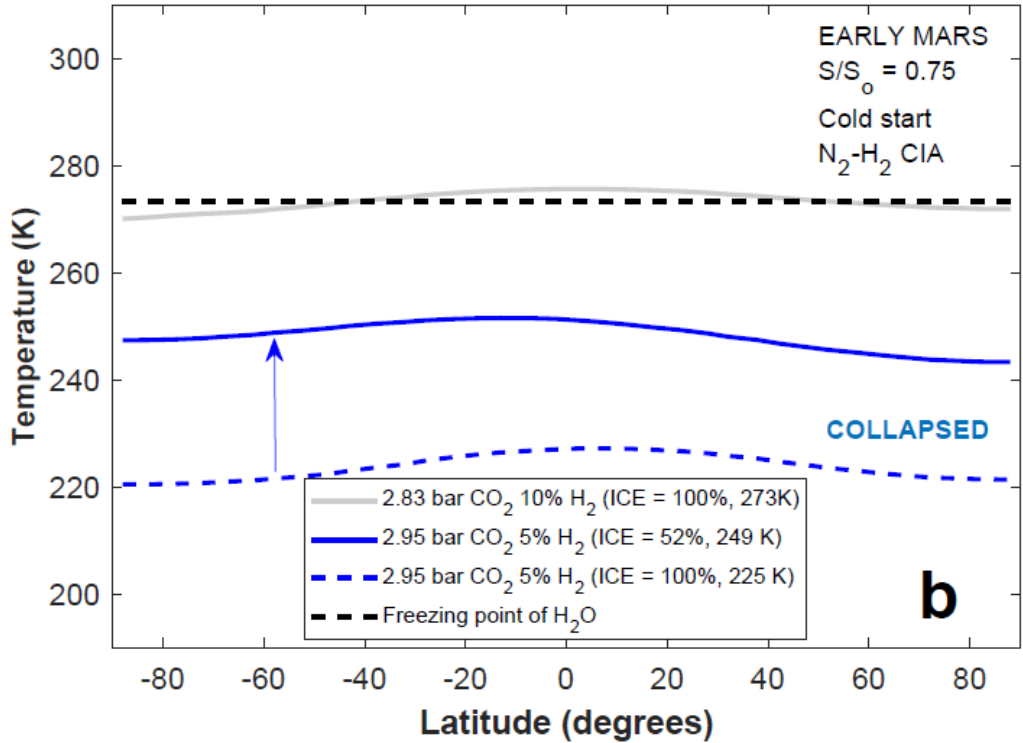

**Figure 6:** Latitudinal mean annual temperature distributions for (a) warm and (b) cold start cases for 1% (red), 5% (blue), and 10% (grey) $H_2$ utilizing $N_2$-$H_2$ CIA as a proxy for $CO_2$-$H_2$ CIA following *Ramirez et al.* [2014a] for various $CO_2$ levels and different land ice fractions. The dashed blue line is the collapsed 5% $H_2$ case.

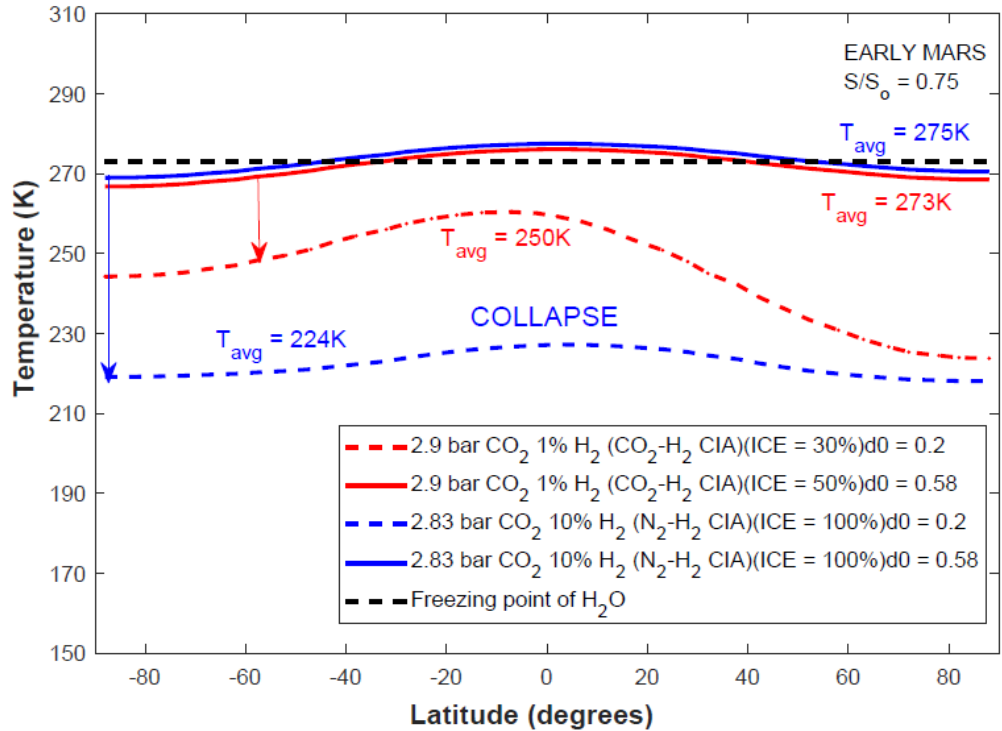

**Figure 7:** Latitudinal mean annual temperature distributions for cold start cases for the (blue) 2.83 bar $CO_2$ 10% $H_2$ $N_2$-$H_2$ CIA case and the (red) 2.95 bar $CO_2$ 1% $H_2$ $CO_2$-$H_2$ CIA case following our previous work [*Ramirez et al.*, 2014a; *Ramirez*, 2017] for different land ice fractions comparing the nominal heat transfer efficiency case ($D_o =$ 0.58; solid red and blue lines) against a low heat transfer efficiency case ($D_o = 0.2$; dashed red and blue lines) that is more appropriate for cold climates.

However, these baseline simulations had assumed that heat transfer efficiency is the same for warm and cold climates. In reality, heat transport efficiency should be lower at cooler temperatures because of inefficient latent heat transport [*Caballero and Langen*, 2005]. We performed limited sensitivity studies at a lower heat transport efficiency ($D_o =$ 0.2) and found that the 2.95 bar $CO_2$ 1% $H_2$ ($CO_2$-$H_2$ CIA proxy) and 2.83 bar $CO_2$ 10% $H_2$ ($N_2$-$H_2$ CIA proxy) cold start cases, which were originally found to yield warm solutions (Figures 5 – 6), become cold at the lower heat transport efficiency (Figure 7). The 2.83 bar $CO_2$ 10% $H_2$ case remained cold, even at relatively low land ice fractions (30%). This again shows that the surface ice problem can be a significant challenge in the deglaciation of planets, even using the CIA of *Wordsworth et al.* [2017]. It also shows that $H_2$ concentrations as high as 10% may not be enough to deglaciate the surface in certain cases.

*3.3 MEBM precipitation and runoff estimates*

We have assessed the precipitation and runoff rates for three ocean sizes and selected latitude bands at 87.5°S and at a representative tropical latitude that has a preponderance of valley networks (27.5°S) [e.g., *Di Achille and Hynek,* 2010] (Figures 8 - 9). In addition to the baseline large ocean of *DiAchille and Hynek* [2010], we considered a smaller ocean from 60° – 90° N (~9% the surface area of Mars) and a medium-sized ocean from ~38° – 90°N (~20% surface area of Mars) [*Batalha et al.*, 2016].

Annually averaged precipitation and runoff rates are similar for both latitudes (Figures 8 – 9). For the 1.85 bar $CO_2$ 5% $H_2$ case, precipitation and runoff rates were ~2.1 and 0.2 mm/day respectively for the large ocean (Figure 8a). Precipitation and runoff rates were ~1.4 and <0.1 mm/day for the medium ocean and ~0.5 and 0 mm/day for the smaller one, respectively (Figures 8b – c). For the 1 bar $CO_2$ 10% $H_2$ case, significant precipitation and runoff were only possible for the large ocean (~0.4 and ~0.05 – 0.1 mm/day; Figure 9a).

This trend is expected if global relative humidity does not vary much from equator to pole. As mentioned earlier, latitudinal gradients on a dense warm early Mars with effective heat transport and relatively flat topography should be rather small (Figure 2). Moreover, although mean surface temperatures are slightly larger near the equator, high seasonal temperatures at the higher latitudes approaching the south pole produce comparatively more precipitation for a short part of the year (Figures 8 - 9). Whereas, precipitation occurs over a larger part of the year at lower latitudes, the most intense seasonal events occur at the higher latitudes. These tend to be competing effects in the model.

Likewise, precipitation and runoff rates for the large ocean should be greater than those for the smaller one. Global precipitation rates for the 1.85 bar $CO_2$ 5% $H_2$ large ocean case were ~1.2 m, which is ~1.6 and ~5 times greater than for the medium and small ocean scenarios, respectively (Figure 8). For comparison, Earth's larger latitudinal temperature gradients produce greater variations in global precipitation patterns (Figure 10). Rainfall rates at a typical tropical latitude in our model Earth average nearly 3 mm/day as compared to only 0.1 mm/day at 87.5°S (Figure 10).

Although precipitation is more continuous at lower latitudes (Figures 8 - 9), the larger seasonal swings at higher latitudes (Figure 2) produce correspondingly greater peaks in precipitation. For the 1.85 bar $CO_2$ 5% $H_2$ large ocean case (Figure 8a), peak rainfall rates are nearly 8 mm/day at 27.5°S, increasing to almost 11 mm/day at 87.5°S. Peak rainfall rates for the medium ocean decrease to ~5 and 7 mm/day at 27.5°S and 87.5°S, respectively. Runoff rates are ~ 0.1 mm/day or less at both latitudes (Figure 8b).

Peak rain fall rates for the small ocean scenario are only ~2 and 2.8 mm/day at 27.5°S and 87.5°S (Figure 8c), respectively, with no appreciable runoff.

For the 1 bar $CO_2$ 10% $H_2$ case (Figure 9), only the large ocean case produces significant precipitation (Figure 9a). Peak rainfall rates are ~1.2 and 1.9 mm/day at 27.5°S and 87.5°S, respectively. In spite of the rainfall, runoff is still very small in this case.

Snowfall does not have a major contribution to the runoff (usually <5%) although it made the largest contribution (~12 – 18%) in the 10% $H_2$ 272 K large ocean case (Figure 9a). With the medium-size ocean, there was almost no rain and no snowfall (Figure 9b). This makes sense. At temperatures well above the freezing point (~280 K), and with a vigorous hydrologic cycle, rain will dominate the precipitation. At temperatures just below the freezing point, snow will make a slight contribution if the precipitation source (ocean) is large enough. Even here, rain dominates global precipitation. As both mean temperatures and precipitation sources decrease, the hydrologic cycle weakens until it shuts down.

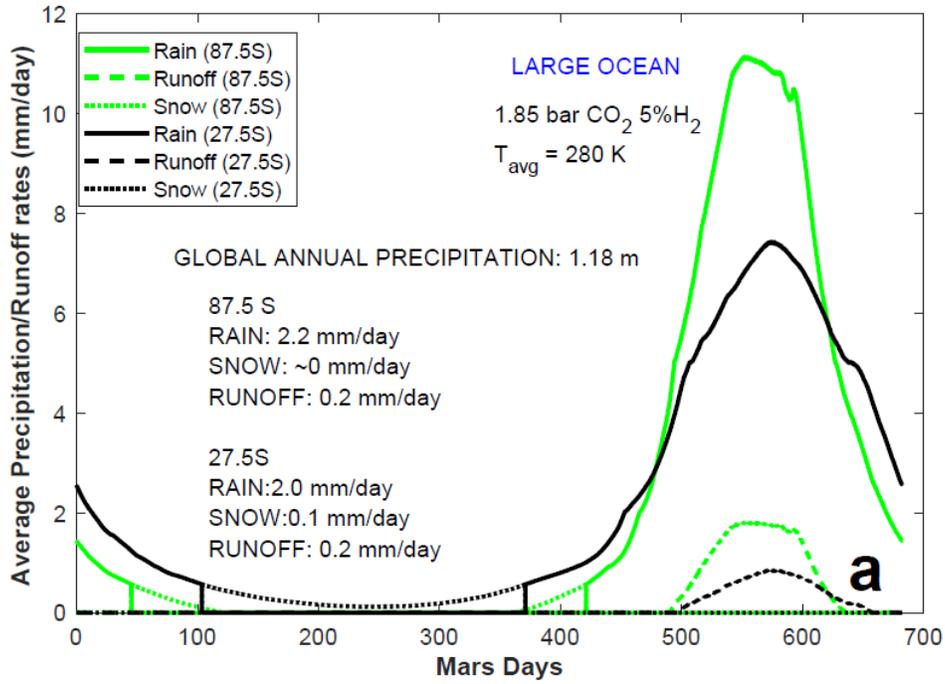
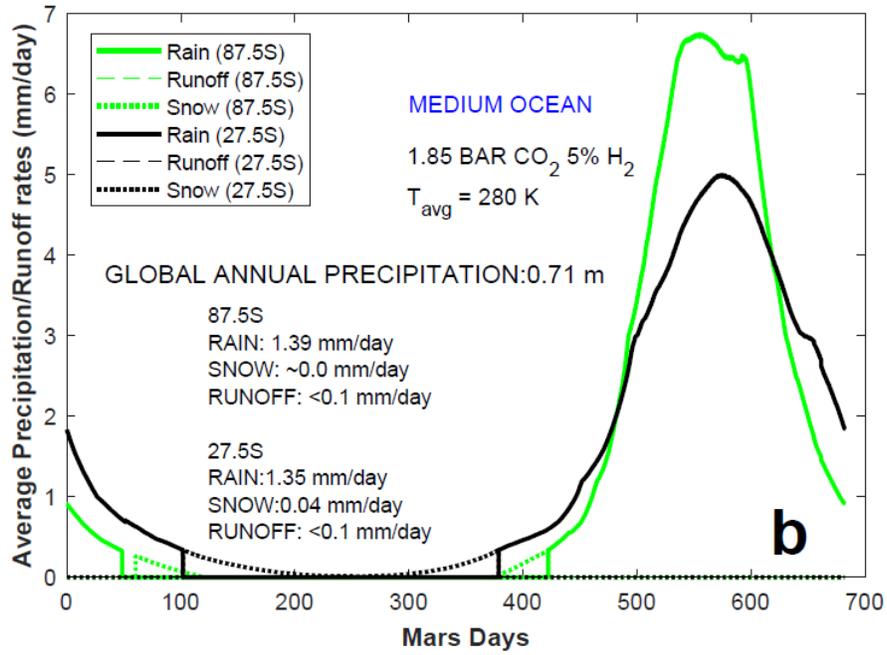

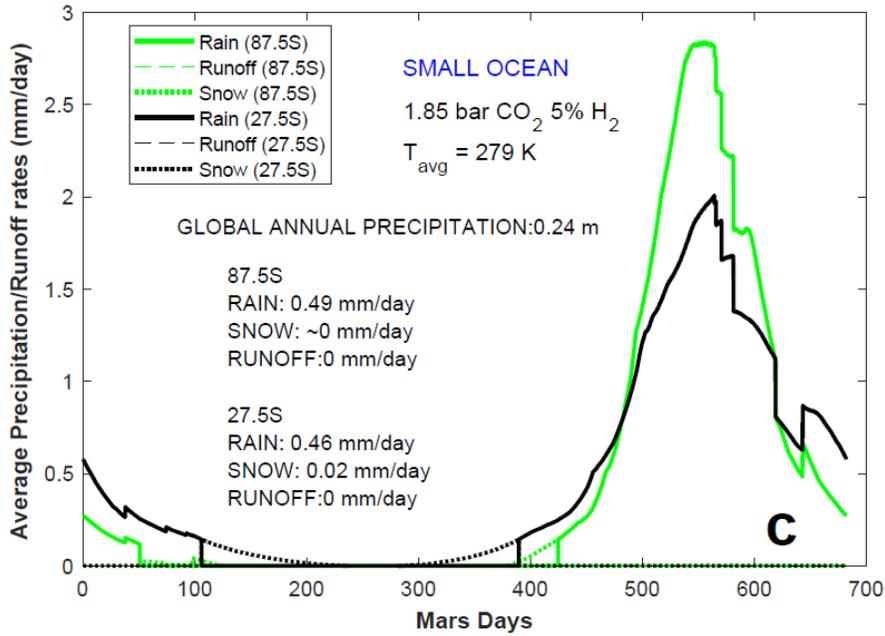

**Figure 8:** Precipitation and runoff rates during the Martian year at (green) 87.5°S and (black) 27.5°S latitudes for a 1.85 bar $CO_2$ 5% $H_2$ atmosphere with (a) large, (b) medium, and (c) small oceans, respectively. Rain , snow, and runoff rates throughout the year are represented by solid, dotted, and dashed curves, respectively. Note that for these and subsequent precipitation plots, the y axis is scaled in a way so that maximum precipitation and runoff rates can be clearly viewed.

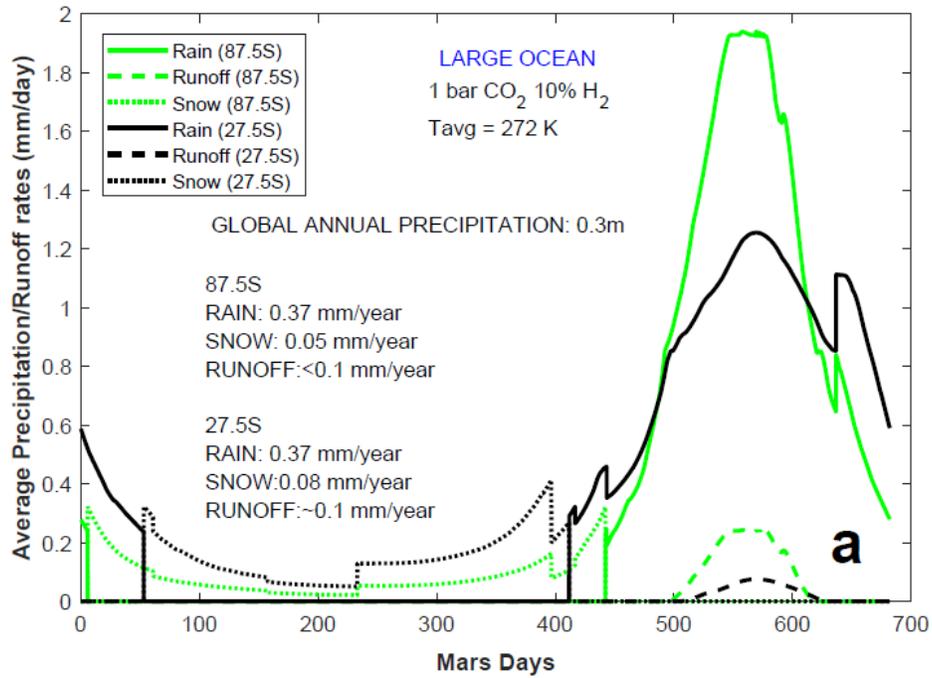
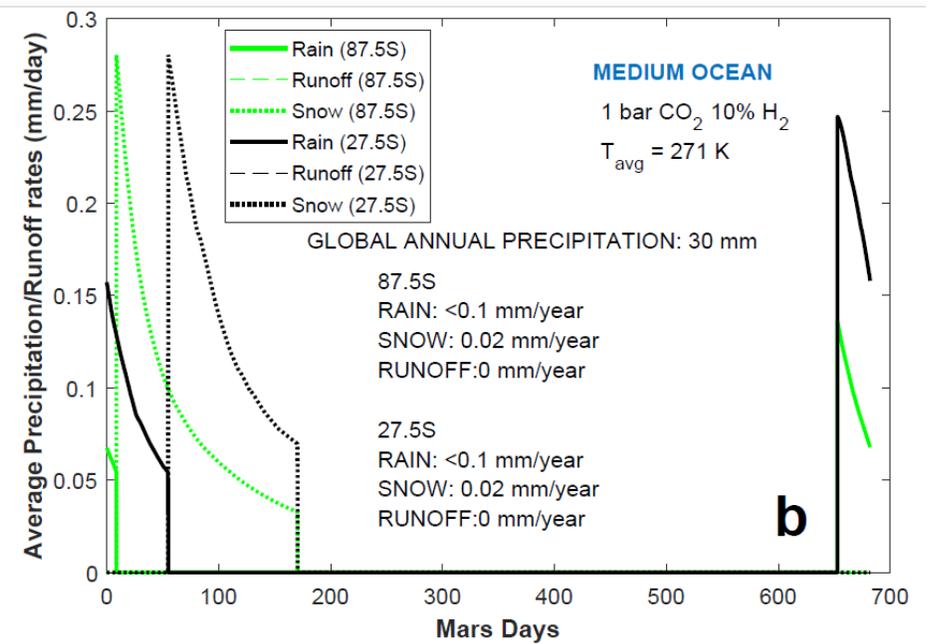

**Figure 9:** Precipitation and runoff rates during the Martian year at (green) 87.5°S and (black) 27.5°S latitudes for a 1 bar $CO_2$ 10% $H_2$ atmosphere with (a) large and (b) medium oceans, respectively. Same color scheme as in Figure 8.

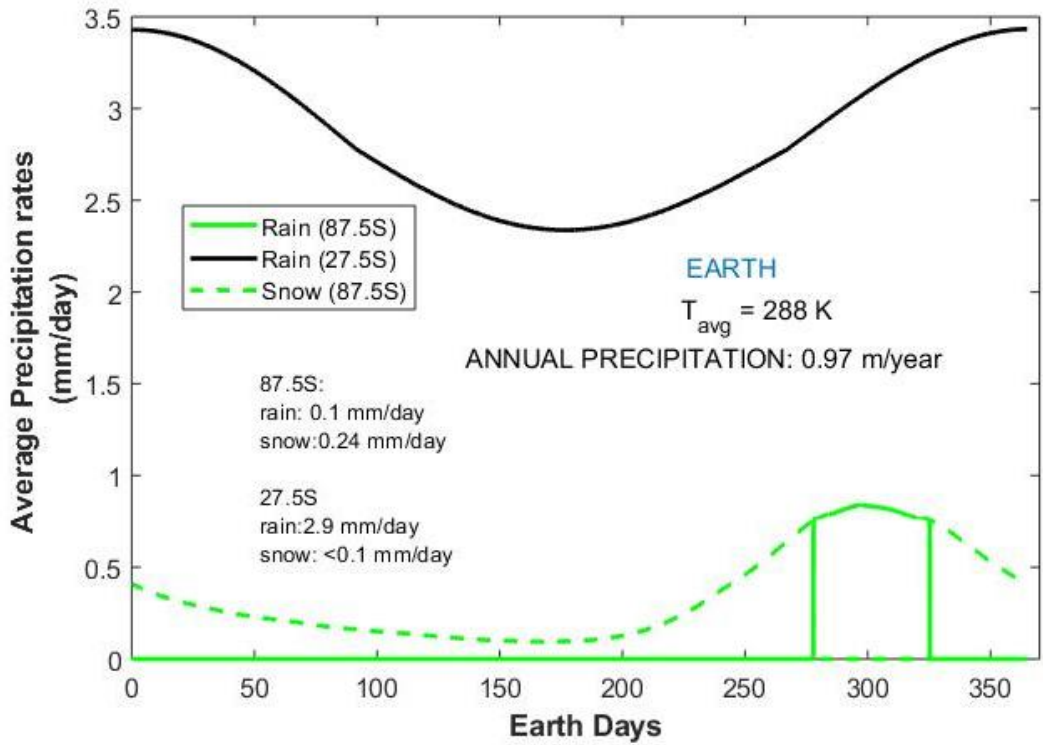

**Figure 10:** Precipitation and snowfall rates during the terrestrial year at 87.5°S (green) and 27.5°S latitudes (black) for the Earth's atmosphere. On Earth, most of the precipitation near the south pole occurs as snow instead of rain. Same color scheme as in Figures 8 and 9.

*3.4 MEBM vs. inferred geologic runoff rates*

The next question is whether MEBM runoff rates for representative climates are consistent with geologic estimates in valley network regions. We have compiled the characteristics of several valley network regions from *Irwin et al.* [2005] and *von Paris et al.* [2015] and computed the inferred mean runoff rates using equation 16 (Table 1). In addition, we have provided peak runoff rates from *Irwin et al.* [2005] and *von Paris et al.* [2015] for comparison. These peak runoff rates were derived from an expression which assumes a 2-year peak discharge recurrence interval for Earth [*Osterkamp and Hedman*, 1982], which *Irwin et al.* [2005] assumed may be representative for early Mars irrespective of what the peak discharge frequency may have actually been. For comparison, our model implicitly assumes that peak runoff occurs once a year (Figures 8 – 9).

MEBM mean runoff rates for the large ocean (~0.1 – 0.2 mm/day) for moderately warm (average mean surface temperature ~280 K) (Figures 8 – 9) $CO_2$-$H_2$ atmospheres are somewhat lower than the inferred mean values from equation 16 (Table 1). However, model peak runoff rates for the large ocean case (~0.1 – 2 mm/day) are consistent with the lower end of the peak runoff rates in Table 1, within a factor of ~2. We suspect that our relatively low modeled runoff rates can be related to the MEBM's inability to model smaller scale intersecting throughflows from surrounding areas.

We also perform sensitivity studies assuming that precipitation on early Mars could have been more intense than what we have estimated in our baseline calculations. This may have been possible if, for instance, atmospheric vertical velocities on early Mars were such that rain clouds could be supported at even lower atmospheric pressures (higher altitudes) than on Earth, effectively increasing $h_q$ [e.g., *Kasting* 1988; *Ramirez and Kasting* 2017] and (possibly) precipitation rates. Precipitation may have also been higher on Mars if storm kinetic energy was higher than on Earth, unlike what we assumed in equation 18. Alternatively, perhaps the precipitation efficiency was higher on early Mars [*von Paris et al.* 2015]. For this calculation, we scale global precipitation (equation 10) up by a factor of 2 and assess the effect on rain, snowfall and runoff rates in this optimistic scenario (Figure 11).

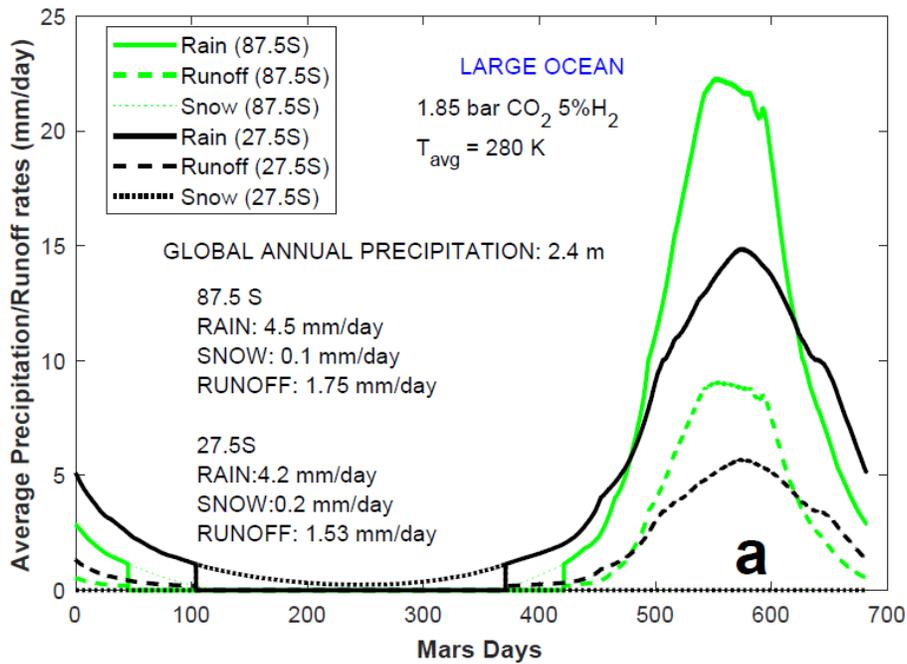

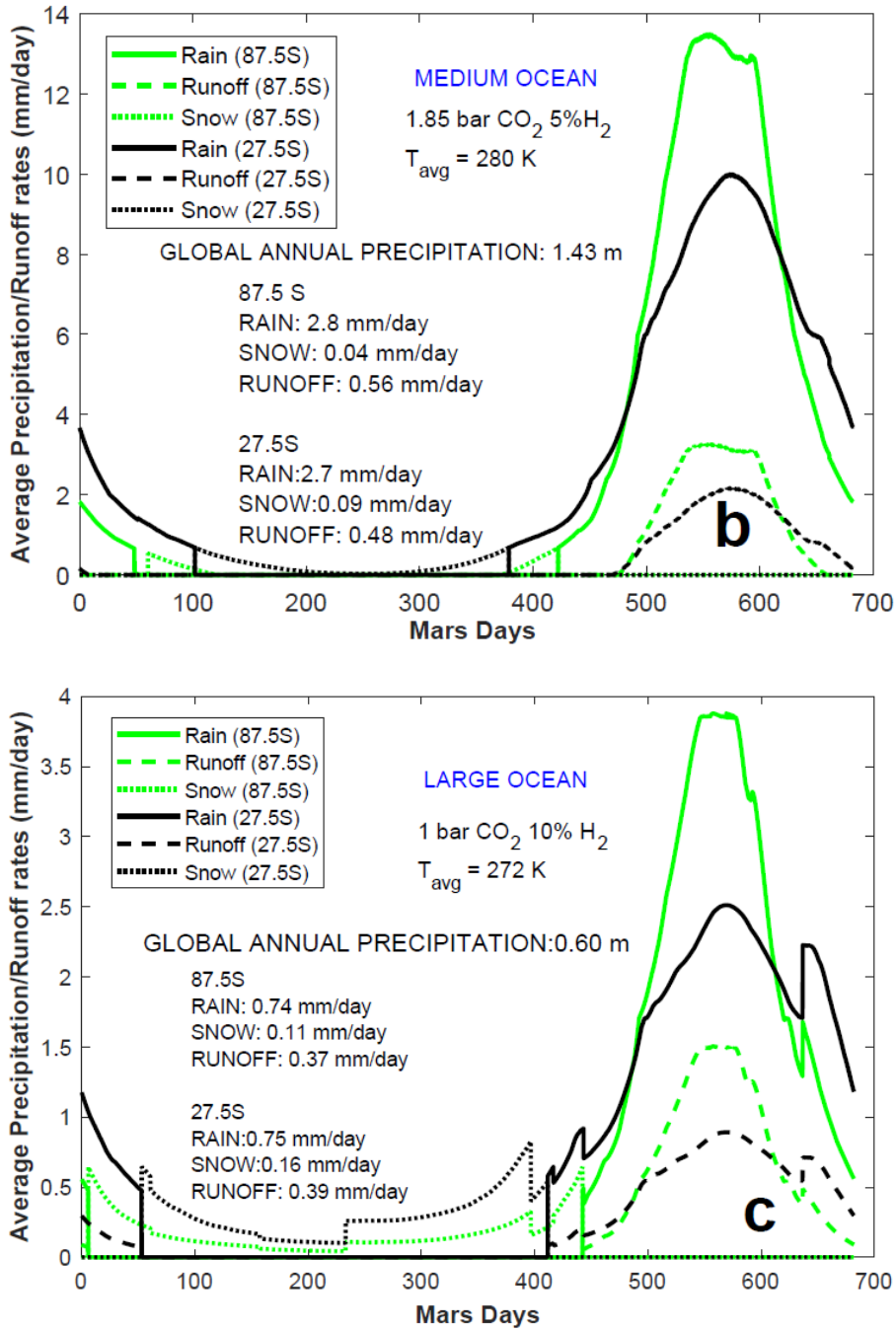

**Figure 11:** Precipitation and runoff rates during the Martian year at the (green) 87.5°S and (black) 27.5°S latitudes, assuming enhanced precipitation (see text), for a 1.85 bar $CO_2$ 5% $H_2$ atmosphere with (a) large and (b) medium oceans, respectively, and the (c) 1 bar $CO_2$ 10% $H_2$ large ocean case. Same color scheme as in Figures 8 – 10.

For the large ocean in the 1.85 bar $CO_2$ 5% $H_2$ case (Figure 11 a), global annual precipitation is 2.4m, as opposed to only ~1.2m in the baseline case (Figure 8a). Rain and runoff rates at both latitudes are ~4.5 mm/day and 1.6 mm/day, respectively. Peak rainfall rates are ~15 – 22 mm/year. For the medium-sized ocean (Figure 11b), global annual precipitation is 1.43 m. At both latitudes, rain falls at ~2.8 mm/day with peak rainfall at ~10 – 13.5 mm/day. Most importantly, the medium-sized ocean in this case, unlike in the baseline scenario (Figure 8b), now has a significant runoff rate of ~0.5 mm/day (Figure 11b).

Likewise, for the 1 bar $CO_2$ 10% $H_2$ large ocean case (Figure 11c), global annual precipitation is ~0.6 m as opposed to 0.3 m in the baseline scenario (Figure 9a). Rain and runoff rates for both latitudes average ~0.75 mm/day and ~0.4 mm/day. Peak rainfall rates of those same latitudes range from ~2.5 – 4 mm/day while peak runoff is ~1 – 1.5 mm/day (Figure 11c). Runoff rates are significantly higher than in the baseline calculation because evaporative losses have a comparatively smaller effect on these larger overland flows (at a given temperature).

The mean runoff rates for these cases (~0.37 – 1.8 mm/day) agree even better with the values in Table 1. The peak runoff rates for the large (~1 - 9 mm/day) and medium (~ 2 -3 mm/day) oceans also coincide well with many of the tabulated values.

Thus, our results suggest that a relatively large northern lowlands ocean in a warmer climate could (in principle) have provided the runoff rates necessary to carve the valleys. However, only by assuming more intense precipitation than in our baseline scenario, does appreciable runoff occur with the medium-sized ocean. In all cases, the computed runoff rates for our smallest ocean are nil and cannot explain valley formation. However, we find that runoff rates of < ~5 cm/year are possible even in the small ocean case if evaporation rates are ~1/5 - 1/3 calculated values (eqn. 12), which requires a significantly lower value for $C_d u$. Although these conditions may be rarely satisfied on a local scale, we do not see this as being a very likely scenario to explain large scale valley formation.

**Table 1: Estimated mean (equation 16) and peak runoff rates**

| Name | Latitude (°) | Longitude (°) | Channel width (m) | Drainage area (m$^2$) | Mean runoff rate (mm/ Mars day) | Peak runoff rate (mm/Mars day) |
|---|---|---|---|---|---|---|
| Parana Valles | 24.06°S | 350.23°E | 180 | 6.2x10$^9$ | 1.85 | 11 |
| Samara Valles | 31.51°S | 347.00°E | 400 | 6.2x10$^{10}$ | 0.72 | 3 |
| Licus Valles | 2.95°S | 126.35°E | 380 | 6x10$^{10}$ | 0.69 | 3 |
| Durius Valles | 17.19°S | 172.09°E | 460 | 1.3x10$^{10}$ | 4.39 | 17 |
| H2539_0000_ND3 | 27°16'S | 128°10'E | 224 | 1.72x10$^9$ | 9.69 | 51.8 |
| H6438_0000_ND3 | 24°54S | 3°26'W | 403 | 1.823x10$^{10}$ | 2.49 | 10 |
| H2081_0000_ND3 | 0°15'N | 124°12'E | 289 | 4.24x10$^9$ | 6.07 | 28.6 |
| H7213_0000_ND3 | 12°21'S | 177°58'W | 285 | 1.412x10$^{10}$ | 1.78 | 8.4 |

*3.5 MEBM computed erosion rates*

Next, we use the USLE (equations 17 and 18) to determine if computed MEBM erosion rates in valley network regions are roughly consistent with geologic erosion estimates. For our high *K* value (0.5), typical precipitation rates in valley network regions are ~0.1 - 2.2 mm/day (Figures 8 - 9), which equate to annual precipitation values in those *latitude* bands of ~7 cm - 150 cm (not to be confused with the *global* precipitation values) .This yields erosion rates of ~2 and 18 kg/m$^2$/year, respectively. Assuming a soil density of ~1520 kg/m$^3$ for Mars [*Hviid et al.* 1997], these rates can be converted to ~0.0013 m/yr and 0.012 m/yr.

At our lower *K* value (0.15), annual erosion rates decrease to ~0.6 and 5.5 kg/m$^2$, respectively, at those latitudes. However, assuming a basaltic rock density of ~3000 kg/m$^3$ for a rock volume fraction of 50%, the soil density is considerably higher ( ~2,260 kg/m$^3$) in the *K* = 0.15 scenario, producing correspondingly lower erosion rates of ~0.0003 m/yr and 0.0024 m/yr. Our calculated erosion rates for large and medium oceans are summarized in Table 2.

**Table 2: Estimated erosion rates in valley network regions for Mars scenarios with different ocean sizes***

| Scenario | Daily precipitation at typical latitude band (in mm) | Erosion rate (kg/m$^2$/Mars year) | Soil erodibility (K) | Erosion rate (m/Mars year) |
|---|---|---|---|---|
| Large ocean (1.85 bar $CO_2$ 5% $H_2$) | 2.1(4.3) | 18(35) | 0.5 | 0.012(0.023) |
| Large ocean (1 bar $CO_2$ 10% $H_2$) | 0.37(0.74) | 4(7) | 0.5 | 0.0027(0.0046) |
| Medium ocean (1.85 bar $CO_2$ 5% $H_2$) | 1.37(2.8) | 12(23) | 0.5 | 0.008(0.015) |
| Large ocean (1.85 bar $CO_2$ 5% $H_2$) | 2.1(4.3) | 5(10) | 0.15 | 0.0023(0.0046) |
| Large ocean (1 bar $CO_2$ 10% $H_2$) | 0.37(0.74) | 1(2.1) | 0.15 | 0.0006(0.0009) |
| Medium ocean (1.85 bar $CO_2$ 5% $H_2$) | 1.37(2.8) | 3.6(6.9) | 0.15 | 0.0016(0.003) |

*Values in parenthesis correspond to the sensitivity study

*3.6. Ice-albedo feedback sensitivity study*

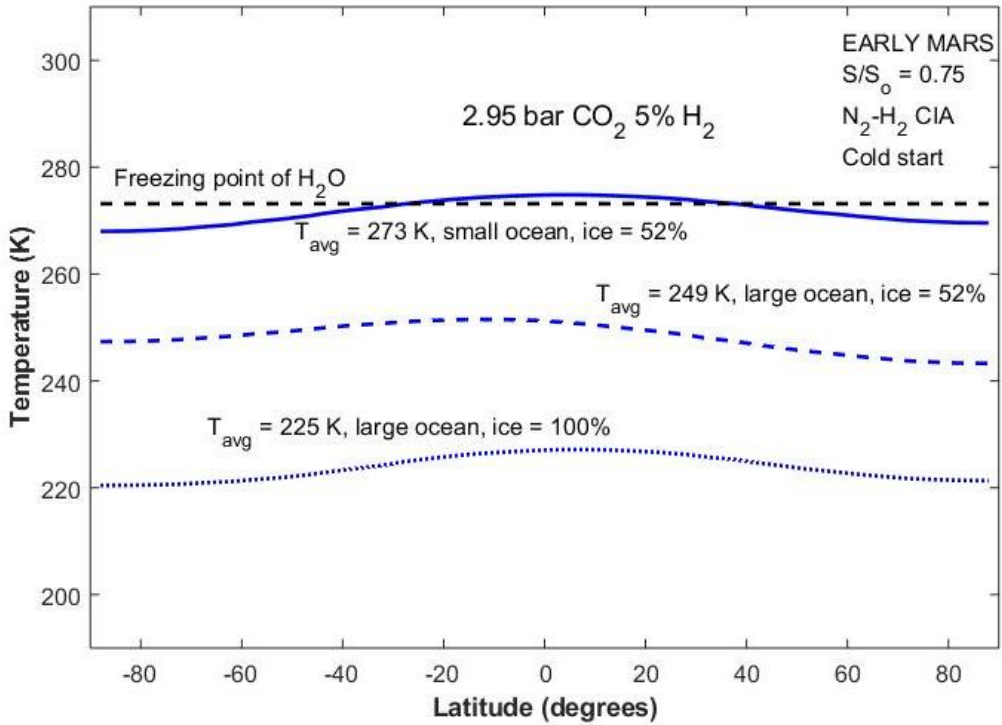

**Figure 12:** Ice-albedo feedback impact on surface temperature for a 2.95 bar $CO_2$ 5% $H_2$ atmosphere assuming the large ocean of *Di Achille and Hynek* [2010] and a smaller ocean (explained in Results). Initial land ice fractions of 52 and 100% are assumed.

We perform limited sensitivity studies on ocean size for a representative cold start atmosphere (2.95 bar $CO_2$ 5% $H_2$) using the $N_2$-$H_2$ CIA proxy assumption (Figure 12). As shown earlier, the cold start scenario yields relatively cold temperatures no matter what the land ice fraction is (Figure 6b; Figure 12). Indeed, if the land ice fraction is 100%, surface temperatures are ~50 degrees below the freezing point (Figure 6b; Figure 12). However, these simulations had assumed a large ocean, which would have a strong ice-albedo feedback at cold temperatures. Relatively warm solutions are possible for this same atmosphere if the ocean is small instead (Figure 12). This is because a much smaller fraction of the land surface is ice in this latter scenario, allowing deglaciation to proceed. Nevertheless, this does not favor transient warming episodes in icy climates that lack a sufficiently large ocean [e,g., *Palumbo et al.*, 2018]. Although the mean temperatures across the planet are "warm" in this case, as we showed above, precipitation and runoff rates with a small ocean are much too small to perform any useful surface erosion (Figure 8).

## 4 Discussion

### 4.1 The ice problem and $CO_2$-$H_2$ CIA

Our results further support the notion that a dense $CO_2$-$H_2$ greenhouse could have explained the early Martian climate [*Ramirez et al.* 2014a; *Ramirez*, 2017]. As we have argued before [*Ramirez,* 2017; *Ramirez and Craddock*, 2018], the high albedo of ice can pose an additional challenge in the deglaciation of an initially cold surface. Deglaciation is not a major issue for surfaces with only modest amounts of ice in many cases, [*Ramirez*, 2017] although this depends on CIA assumptions and heat transport efficiency (Figure 7). In a pure $CO_2$-$H_2O$ atmosphere on early Mars, surface temperatures can be some tens of degrees cooler once this ice-albedo feedback is included (Figure 5a). However, the additional warming by hydrogen can diminish this effect.

With the $N_2$-$H_2$ CIA assumption for $CO_2$-$H_2$ [*Ramirez et al.* 2014a], our model predicts that a warm early Mars solution from a cold start is possible at hydrogen concentrations exceeding ~5%, irrespective of atmospheric composition (Figure 5b). In contrast, warm solutions starting from a cold initial condition are possible with the *Wordsworth et al.* [2017] $CO_2$-$H_2$ CIA for $H_2$ concentrations equal to or exceeding 1% (but see Figure 7 for exceptions). Thus, warm solutions from cold starts can be achieved with either CIA assumption given high enough hydrogen concentrations and land ice fractions that are not significantly higher than ~ 0.5 and ~0.9 with the $N_2$-$H_2$ and $CO_2$-$H_2$ CIA assumptions, respectively (Figures 5 -7). That said, the exact greenhouse gas pressures and concentrations required differ greatly among the two sets of assumptions.

Thus, transient warming become difficult for mechanisms that require repeated freezing and thawing from surfaces with high ice coverage (e.g., limit cycles)[*Batalha et al.,* 2016], especially since the albedo of $CO_2$ ice is relatively high (see Methods). For the icy highlands hypothesis [e.g., *Wordsworth et al.* 2013], typical surface ice fractions appear to be ~ 25 – 30%. Thus, the ice problem would ostensibly not be a major issue in most cases (although deglaciation may still be difficult to achieve even in this scenario in some cases; Figure 7). However, even beyond this consideration, is this even enough ice to produce the snowmelt required to form all of the valleys and other erosional features in a series of transient warming events? This question has yet to be answered, but it is a crucial one for the survival of the icy highlands hypothesis, because if ice cover and amounts fall below some threshold, it may not be possible to produce sufficient snowmelt to form the observed fluvial features.

Furthermore, the *Wordsworth et al.* [2013] and *Forget et al*. [2013] radiative transfer model predicts significant warming from $CO_2$ ice clouds while more accurate schemes have subsequently shown that such clouds should exhibit little (if any) warming [e.g., *Kitzmann* 2016]. If the latter study is correct, that would imply significantly lower surface temperatures and higher amounts of surface ice (plus possibly $CO_2$ ice) than predicted in the baseline icy highlands hypothesis. This follows from our understanding of the ice-albedo feedback for snowball climates [e.g., *Pollard and Kasting*, 2005]. Such a scenario could greatly increase the difficulty of transient warming via melting of icy highlands sheets, although the strength of the feedback depends on unknown factors, including water inventory and the global topography. It is also possible that after a transient warming episode, $CO_2$ get sequestered within the surface water ice due to the density contrast, reducing the amount of atmospheric $CO_2$ available for subsequent warming episodes [*Turbet et al.,* 2017].

Our calculations have ignored other potential mechanisms, like dust fall from periodic dust storms, which can temporarily lower the surface albedo slightly on present Mars in certain locations [e.g., *Fenton et al.,* 2007]. Such storms can also warm the planet by increasing the atmospheric opacity and decreasing the planetary albedo [*Forget et al.,* 2013]. Whether similar storms occurred on early Mars is unknown but their impact on surface temperatures seems to be no more than ~ 1 - 10 K or so in the most optimistic scenarios, according to one model [*Forget et al.,* 2013]. The impact of dust fall would be even less at very low temperatures [~210 K, Figure 5a], especially if the greenhouse effect of $CO_2$ ice clouds had been overestimated [*Kitzmann et al.,* 2016], This is because the greenhouse effect is very weak at those temperatures. For comparison, dust storms on Mars today increase global surface temperatures by < 1 K [*Fenton et al.,* 2005]. Thus, dust storms are not likely to trigger deglaciation unless the atmosphere is already on the brink of deglaciation.

Although the *Turbet et al*. [2019] CIA have large error bars, their work suggests that the real $CO_2$-$H_2$ CIA strength may be near the midpoint of the two sets of CIA assumptions used. This would mean that atmospheric pressures for warm conditions would exceed 2 bar for our ~1 – 5% $H_2$ cases considered, which is higher than those for current estimates of paleopressure [e.g., *Kite et al.,* 2014; *Hu et al.,* 2015]. However, at still higher $H_2$ concentrations, paleopressures lower than 2 bar might be possible [e.g., *Ramirez*, 2017]. Nevertheless, further progress requires the determination of more accurate $CO_2$-$H_2$ CIA cross-sections.

*4.2 The lack of glacial features in valley terrains and the warm vs. cold hypothesis*

An early Mars that was relatively warm is not only supported by these climate modeling simulations, but by the geologic record itself [*Ramirez and Craddock*, 2018]. The absence of glacial features, including eskers, kames, and frost wedges in ancient valley terrains all suggest that the valleys were not formed in a cold and icy climate. Although a recent study presents evidence of possible ice-related features located in the highlands region south of Terra Sabae [*Bouquety et al.* 2019], such features are located at higher elevations far away from the valley networks themselves. Moreover, no periglacial features have been identified in this area and so there is no clear connection between the U-shaped valleys (which the authors inferred to be glacial in origin) and the valley networks. Thus, there remains no evidence of glacial features in valley network terrains. Our simulations are also consistent with such a scenario.

*Wordsworth et al.* [2015] argue that the lack of glacial features in geologic terrains not only poses problems for cold and icy scenarios, but for the warm and wet case, because a frozen ocean after the end of the warm period would have left behind evidence of wet-based glaciation, which is not observed. However, this partially depends on the exact ocean size by the end of the warm period, atmospheric temperatures, and on the geothermal heat flux at the time. If we assume that the ocean froze following the warm period, we can calculate the maximum glacial ice thickness before the base melts and wet-based glaciation is triggered ($z_{max}$) as [e.g., *Pollard and Kasting,* 2005] (equation 19):

$$z_{max} = k \frac{\Delta T}{F_g} \tag{19}$$

Here, $k$ is the thermal conductivity of water ice (~2.5 Wm$^{-1}$K$^{-1}$), $\Delta T$ is the temperature difference between the top of the glacier and the melting temperature of H$_2$O ice at the glacial base, and $F_g$ is the geothermal heat flux. For a hard snowball Earth model, the melting temperature of sea ice is ~ 271 K whereas the near-surface atmospheric temperature is ~ 246 K [*Hyde et al.,* 2000], yielding $\Delta T$ = 25 K. Thus, $z_{max}$ = ~ 1 km. However, surface temperatures on a cold early Mars postdating the warm period, would have likely been much lower than on a snowball Earth due to the reduced insolation, perhaps ~210 K or even less, as our cold solutions show here (Figure 5a). This could yield $\Delta T$ values in excess of ~60 K. If we assume that the geothermal heat flux of early Mars was the same as for the Earth (~0.06 Wm$^{-2}$) [*Montesi and Zuber*, 2003], $z_{max}$ could easily have been a few km, and depending on the ocean size, could possibly explain the lack of wet-based glaciation. If the geothermal heat flux was lower than present Earth's, which might be reasonable as the planet gradually lost its ability to retain its internal heat over time [e.g., *Kasting* 1988], $z_{max}$ could have been even larger still. It is also possible that the ocean started relatively large and shrunk over time as hydrogen hydrodynamically escaped to space [e.g., *Kurokawa et al.,* 2014; *Villanueva et al.,* 2015], with any remaining water becoming sequestered within the subsurface [e.g., *Mouginot et al.,* 2012; *Usui et al.,* 2015; *Wade et al.,* 2017].

Nevertheless, the issue with a cold and icy Mars is not just the lack of evidence for wet-based glaciation, but rather the lack of apparent glaciation of any sort in valley terrains [e.g., *Davis et al.,* 2016; *Ramirez and Craddock*, 2018]. For instance, dry-based glaciation features (e.g. frost wedges) are not found at valley network terrains either [e.g., *Grotzinger et al.,* 2015], which would form even if the glaciers were too thin for wet-based glaciation to occur. In contrast, abundant evidence of dry-based glaciation exists throughout Mars history following the period of valley formation [e.g., *Head and Marchant*, 2003; *Dickson et al.,* 2008]. Thus, it may be reasonable to expect periglacial features associated with a Noachian ocean, assuming it had existed. However, unlike the relatively better preserved outcrops throughout the southern highlands, some of which we have been able to closely observe with rovers and question the idea of a very icy early climate [e.g., *Grotzinger et al.,* 2015; *Davis et al.,* 2016; 2019; *Ramirez and Craddock*, 2018], the same cannot be done to falsify a possible Noachian ocean. This is because the northern lowlands largely represent later Hesperian and Amazonian lava flows [e.g., *Tanaka et al.,* 2005; 2014] that would have subsequently covered and destroyed any periglacial features that may have existed following the disappearance of this early ocean.

### *4.3 Runoff rates in warm versus cold and icy climates*

Our study is the first to use climate models to corroborate independent estimates of inferred geologic runoff rates for valley network regions. Recent analyses suggest that the early Martian climate may have been arid to semi-arid [e.g., *Craddock and Howard,* 2002; *Barnhardt et al.,* 2009; *Hoke and Hynek*, 2009; *Ramirez*, 2017; *Ramirez and Craddock*, 2018; *Davis et al.,* 2019). Wetter climates than this would produce greater fluvial dissection in valley networks than what is actually observed [e.g., *Ramirez and Craddock*, 2018]. Some rough geographical definitions define arid and semi-arid climates on Earth as those having less than 20 cm and between 20 – 50 cm of annual precipitation, respectively [*Grove* 1977]. Scaled to the Martian year, these might correspond to an upper limit of ~40 cm for arid climates and between 40 – 100 cm/year for semi-arid ones for Mars. If we utilize these (admittedly crude) definitions, then the 280 K 1.85 bar $CO_2$ 5% $H_2$ large ocean case (Figure 8a) is only slightly wetter than a semi-arid climate, requiring either a slightly higher cloud cover (Figure S2b) or slightly lower relative humidity than what we have assumed (see section 4.8). For comparison, the medium ocean case satisfies our criterion for semi-arid climates although runoff rates are lower than in other scenarios (Figure 8b). The small ocean case (Figure 8c) satisfies the criterion for an arid climate, although we find that runoff rates in that case are zero unless unusually low evaporation rates are assumed (see Results). If precipitation was even higher on early Mars than in our baseline calculations, the 272 K 1 bar $CO_2$ 10% $H_2$ large ocean scenario would also satisfy this suggestion (Figure 11c).Thus, mean surface temperatures near or slightly above the freezing point of water (~270 – 280 K) appear to be required to explain the observed runoff in semi-arid climates. This result is relatively insensitive to both the $CO_2$-$H_2$ CIA assumptions utilized or the exact atmospheric composition.  This also implies that the seasonal melting scenarios with sub-freezing climates proposed by *Palumbo et al.,* [2018] would not produce sufficient water over time to explain observations. Seasonal snowmelt in an icy climate would not be able to produce the required runoff rates to form the valleys either [*Kite et al*., 2013]. A new study invokes the situation in Earth's Antarctica Dry Valleys to suggest that seasonal warming on early Mars may have occurred with a ~255 K mean surface temperature [*Kite et al.,* 2020]. However, our model predicts negligible hydrologic activity at such low mean surface temperatures**.** As we have shown, seasonal snowmelt in a much warmer climate (with a small or no ocean**)** would *also* not produce the required runoff because the water source is not large enough (Figures 8 – 9). This poses a real limitation for transient warming mechanisms in icy climates, because integrated over time, they should produce even less runoff than the continuously warm scenarios we explore here.

Rain was not only required but it would have been much more abundant than snowfall, even at mean annual temperatures slightly below the freezing point (Figures 8 – 11, S2). At still lower temperatures, the planet freezes and neither rain nor snow become available. Although, higher runoff rates at lower mean annual surface temperatures are possible with highly eccentric (e = 0.17) orbits [*Palumbo et al.*, 2018], such scenarios are not supported by our current understanding of Mars' orbital evolution [*Laskar et al.*, 2004]. Overall, our results are consistent with the idea that rainfall was a major global process impacting the observed erosion on Mars [*Craddock and Howard*, 2002; *Ramirez and Craddock*, 2018]. They are also consistent with a warm climate that could have been sustained by a relatively large ocean in the late Noachian-early Hesperian, similar to the one proposed in *Di Achille and Hynek* [2010], although a somewhat smaller ocean may have been possible also (Figures 8 – 11, S4).

Our results also show that the presence of such a northern lowlands ocean does not suggest very warm and wet climates akin to Earth's tropical regions. Contrary to the arguments of *Wordsworth* [2016], as we have shown, a semi-arid climate is possible (even expected) if a northern lowlands ocean had existed on a warm early Mars (Figures 8 – 11, S2, S4). A key reason for this is that a warm climate with a 273 K mean surface temperature will be considerably drier than one with an "Earth-like" 288 K. This can be seen through the exponential dependence of saturation vapor pressure on temperature according to the Clausius Clapeyron equation. Even though the temperature difference seems relatively slight, the $H_2O$ saturation vapor pressure at 288 K is already ~3 times greater than at 273 K. Thus, a planetary atmosphere with a mean surface temperature of 273 K should be considerably drier than the Earth, even assuming all other characteristics are equal (including water inventory). Moreover, the ratio of ocean to land area in our early Mars simulations is also smaller in comparison to the Earth, which also suggests a drier hydrologic cycle. However, this is more complicated on a warm early Mars because the lower gravity suggests a thicker atmospheric water vapor thickness ($h_q$) than the Earth at the same surface temperature and relative humidity, leading to relatively high precipitation rates (equation 10). This contributes to why surface temperatures exceeding ~280 K produce climates that are too wet for valley formation in this model.

*4.4 Estimates of late Noachian-early Hesperian erosion rates*

Geologic estimates of *average* Noachian erosion rates are ~0.1 – 10 µm/year [e.g., *Golombek and Bridges*, 2000], with a best estimate of 7.7 µm/year (~15 µm per Mars year) for the late Noachian ~ 3.83 – 3.7 Ga, for a total erosion of ~1,000 m [*Golombek et al.*, 2006]. This best estimate is at least an order of magnitude lower than our erosion rate estimates in Table 2. We emphasize that this best estimate is merely an average rate over the entire timeframe (~130 million years), as erosion rates could potentially have been much higher than this mean value over much shorter periods of relatively higher fluvial activity. Estimates for the total erosion in the late Noachian to early Hesperian were not given, but total erosion between the Middle Noachian and Hesperian was estimated at ~300 – 2,300 m. [*Golombek et al.,* 2006]. If we estimate that the total erosion during the late Noachian – early Hesperian was ~1,500 m, then it would only take ~123,000 – 1,050,000 years (~65,000 – 560,000 Mars years) to explain the observed erosion if $K$ = 0.5. In comparison, if $K$ is 0.15, these rates modestly increase to ~940,000 – 4,700,000 years (~500,000 – 2,500, 000 Mars years). These numbers assume a continuously warm climate with steady rainfall (with seasonally higher periods of precipitation, see Figures 8 – 11; Table 2). However, there are some caveats with such assessments. The regolith was assumed to compose largely of relatively erodible soils, but this need not have been the case. A higher bedrock fraction than we estimated could decrease erosion rates by an order of magnitude or more [e.g., *Kite et al.* 2014]. Erosion rates would also decrease if soils were more permeable than estimated, further decreasing $K$. If $LS$ was smaller than assumed, erosion rates would decrease even further.

Overall, our numbers are consistent with some recent estimates suggesting relatively short valley formation timescales of ~$10^4$ – $10^7$ years [e.g., *Howard* 2007; *Barnhardt et al.,* 2009; *Hoke et al.,* 2011; *Orofino et al.,* 2018]. However, the biggest assumption is that the current surface geologic expression represents all the erosion that ever took place. This need not have been the case though. Some of the evidence for erosion could have been removed through subsequent impacts or mantling [e.g., *Kreslavsky and Head,* 2002]. Most of the quantified evidence for erosion is based on valley network morphology [e.g., *Luo et al.,* 2017]; however, there was also widespread crater modification during the Noachian [e.g., *Craddock et al.,* 1997] as well as complete eradication of smaller impact craters and redistribution of higher frequency topography [*Cawley and Irwin,* 2018], and the total erosion from these processes has not been adequately quantified. Thus, estimates of erosion rates and valley formation timescales remain very poorly-constrained.

*4. 5 Global precipitation and valley network locations*

Our model predicts that valley networks should have formed all throughout the southern hemisphere. Our sensitivity studies on heat transport efficiency in warm climates further support this notion (Figure S5). Our simulations are consistent with the warm simulations of *Wordsworth et al*. [2015], whom also obtain significant precipitation from equator to pole in spite of different model assumptions regarding topography, atmospheric-ocean heat transfer, and the ice-albedo feedback (further discussed in the Introduction). However, observed valleys are primarily located between 0 and 60°S, with a higher concentration between ~0 and 30°S [*Hynek and Hoke,* 2009]. Relatively few networks are found near the south pole. So, if precipitation was global, why are there not more observed valleys between 60 and 90°S? Observations suggest that thick mantling and icy deposits cover the landscape at latitudes poleward of ~60°S, obscuring or destroying any evidence of valley formation at those locations [e.g., *Kreslavsky and Head*, 2002]. Thus, the current valley network distribution is not representative of all of the valleys that had existed [*Ramirez and Craddock*, 2018]. Unfortunately, this is why using climate models to compare against the current valley network locations, as has been recently attempted [Bouley et al. 2016; *Wordsworth et al.*, 2015], is not effective in validating such models. What climate models can do is attempt to predict where the valleys *may* have formed preceding the erosional and depositional processes that have forever erased them from the rock record. Nevertheless, the fact that there are some remnant networks between 60°S and 90°S [*Hynek and Hoke*, 2009] hints that the distribution could have very well been global. Also, valley network density distributions could be further affected by topographical variations (see section 4.7).

*4.6 Valley network formation and mineral formation at Mawrth Vallis and similar locations*

The analysis here also has interesting geochemical implications. A recent study suggests that relatively rapid timescales (< ~ tens of thousands of years) and local surface temperatures well above 300 K were necessary to form the abundance of clay minerals observed at Mawrth Vallis, Nili Fossae, Gale crater and similar locales on ancient Martian terrains [*Bishop et al.*, 2018]. These authors argued that such high surface temperatures were necessary to obtain reaction rates that were great enough to produce the surface minerals. On Earth, mean average temperatures near the equator can exceed 300 K although the mean surface temperature is significantly lower (288 K). However, caution should be exercised in extrapolating the Earth to Mars. As our results show here (Figures 2 – 4), the equator-pole temperature gradient in a warm early Martian atmosphere would be significantly lower than that for the Earth, leading to a closer correspondence between local and mean surface temperature. In other words, to achieve local mean surface temperatures above 300 K, *global* mean surface temperatures would certainly also have to exceed 300 K for the greenhouse gas solution to explain such clay mineral formation. The resulting climates would require atmospheric pressures well above 3 – 5 bar or even more [*Ramirez*, 2017], which apparently exceed current estimates of paleopressure [e.g., *Kite et al.,* 2014; *Hu et al.,* 2015; *Craddock and Lorenz*, 2017; *Kurokawa et al.,* 2018]. Moreover, the results here do not support mean surface temperatures being much higher than ~280 K or perhaps valley network runoff rates would be even higher than geologic estimates (Table 1). That said, *seasonal* maximum temperatures can exceed 300 K for our initially warm 1.85 bar $CO_2$ 5% $H_2$ (2 bar) atmosphere for latitudes below 15°S (Figure 2b), but these latitudes do not coincide with the higher latitude locations of Mawrth Vallis, Nili Fossae, or Gale crater (which are located closer to the purported ocean shoreline) although such seasonal warming could possibly contribute to clay formation at lower latitudes. The presence of a large northern ocean also implies that temperatures above 300 K are difficult to achieve because of the small seasonal variations at those equatorial latitudes (Figure 1).

Unless it is later found that such clays can form at lower temperatures (~ 270 K - 280 K), the above arguments imply either that the paleopressure estimates are incorrect, and the early atmosphere was even thicker than imagined, or another unrelated mechanism was responsible for clay formation at places like Mawrth Vallis. The fact that many of these deposits seem to predate the valley networks, plus given that their distribution does not seem to coincide very well with the valley locations [*Bishop et al.,* 2019], suggests that these clays were formed via different mechanisms [e.g., impacts, *Haberle et al.,* 2019] than what was responsible for valley network formation.

*4.7 The flat topography assumption*

A big assumption in this work is that of a flat topography for early Mars. As mentioned before, this is based on geologic evidence and geophysical models suggesting that the bulk of Tharsis was not in place at the start of valley network formation [e.g., *Craddock and Greeley*, 2009; *Citron et al.,* 2018]. The obvious problem with this assumption is that early Mars was not flat, only flatter than present. On Earth, orographic precipitation on tall mountains enhances precipitation on windward slopes and a reduction in condensation on leeward slopes. According to *Roe* [2005], orographic effects on Earth exert significant control over precipitation distribution patterns over mountain ranges with elevations exceeding ~1.5 km. At a similar surface temperature, and assuming this height can be safely scaled by the water scale height for a planet with Mars' gravity, this may be equal to (1.5 x 9.81/3.73) ~4 km on early Mars. The maps in *Citron et al*. [2018] for an early Mars with a northern ocean suggest an elevation range of ~ -2 to ~6 km in the southern highlands. However, most of the northern lowlands in *Citron et al*. [2018] have an elevation below 4 km, with zonally-averaged values being under ~3 km, which, again, seems to validate our assumption (to first order). That said, our model likely overestimates precipitation in regions where elevations exceed ~4 km, although by how much is unclear. However, the presence of inverted channels in Arabia Terra [*Davies et al.,* 2016; 2019], which are located on the leeward side of Tharsis, suggests that the rain shadow effect predicted by previous GCM simulations [*Wordsworth et al.,* 2015] may have been overestimated. Either that, or the Tharsis region was not nearly as high during valley network formation as *Wordsworth et al.* [2015] thought. In either case, we suggest that our flat topography assumption is a good point of comparison with future studies. Assuming precipitation had occurred on some leeward slopes, that may just mean that valley formation in those cases took somewhat longer than what we estimate here. Few studies on orographic precipitation on early Mars exist [e.g., *Scanlon et al.,* 2013; *Wordsworth et al.,* 2015], but predicting the magnitude and location of this effect remains challenging for models, even when the topography and weather patterns are known, like they are for present Earth [*Roe,* 2005].

*4.8 The constant relative humidity assumption*

As with similar 1-D and 2-D calculations, one of our biggest assumptions is the use of a constant relative humidity (RH) profile. Although the exact prescription for RH does not have a big effect on calculated mean surface temperatures for these dense $CO_2$- $H_2$ atmospheres [e.g., *Ramirez et al.,* 2014a], it could influence computed precipitation and runoff rates across the planet. On Earth, seasonal variations in RH from one hemisphere to the other vary by up to ~10%, though usually less [*Peixoto and Oort*, 1996]. Polar RH remains relatively high on Earth because of the low saturation vapor pressure. Near the equator, RH on our planet remains high because of increased evaporation at higher temperatures. These effects approximately cancel out and keep RH differences with latitude to a minimum. According to equation 10, a 10% decrease in surface RH from 0.77 to 0.69 (assuming all else equal), decreases precipitation rates within a given latitude band to $(0.69/0.77)^3$ ~72% of calculated values, which to first order, is roughly comparable to our baseline calculated values. Nevertheless, we predict that latitudinal gradients in RH for these dense and warm early Martian atmospheres should not be this high because of the relative homogeneity in mean surface temperatures (Figure 2), particularly with a reduced Tharsis region. If true, that would further justify our constant global surface RH assumption.

The other concern is whether *global* average surface RH on a warm early Mars could have been as high as 0.77. Using the same RH = 0.69 value as above but applied globally, our model predicts virtually no runoff in the 1.85 bar $CO_2$ 5% $H_2$ large ocean case (Figure 8a) unless precipitation efficiency was somehow higher on early Mars (Figure 11) or global surface temperature was higher because cloud cover was low (Figures S1 – S4 ). However, such a dry early Mars is inconsistent with the abundant fluvial evidence and so it is likely that a warm early Mars did not have such a low average surface RH, at least in areas of valley network formation. Of course, if average global surface RH was higher than 0.77, higher precipitation and runoff rates than what we calculated here would be possible.

*4.9 The importance of atmospheric-ocean heat transport*

As has been shown, accurately modeling the influence of oceans on global heat transfer is crucial for simulating warm planetary climates [e.g., *Cullum et al*., 2014]. However, most 3-D models simulating warm climates do not dynamically couple the atmospheric and ocean circulations, and instead assume a static heat flux into the ocean [e.g., *Shields et al.,* 2013; *Wordsworth et al.,* 2015; *Bin et al.,* 2018]. This is partially due to the enormous computational expense that arises from properly incorporating dynamic oceans [*e.g., Ramirez,* 2018], but also from uncertainties in the implementation of atmospheric-ocean heat transfer on planets that are different from the Earth. This is a key point because a lively debate exists in the terrestrial literature regarding the relative importance of ocean and atmospheric heat transport, including how these impact cloud cover, relative humidity, and planetary circulation patterns [e.g., *Held* 2001; *Czaja and Marshall,* 2006; *Rose and Marshall,* 2009; *Koll and Abbot,* 2013; *Rose,* 2015]. However, it remains unclear how well these 3-D models derived for a planet with a global ocean, scale to a much smaller planet with a smaller ocean coverage and a completely different atmosphere, like a warm early Mars.

We have recently argued that more observations of the climate and weather patterns on other planets, including exoplanets, are essential for improving our understanding of planetary climates [*Ramirez et al*., 2018; *Ramirez,* 2018]. We have also argued that using simpler and faster models (e.g., EBMs, radiative-convective climate models) will continue to be useful for parameter space exploration, including investigating scenarios that are relatively difficult to assess with more complex models.

**5 Conclusions**

Using an advanced energy balance model, we find that the observed surface erosion, plus valley network formation, require mean annual surface temperatures near or slightly above the freezing point of water. In contrast to previous work [e.g., *Wordsworth*, 2016], the presence of a northern lowlands ocean during the late Noachian-early Hesperian does not necessarily lead to very moist and warm "Earth-like" climates, and is instead consistent with the semi-arid to arid conditions that may have typified early Mars. This "warm and dry" state, which *Wordsworth* [2016] deemed difficult to achieve, is most consistent with the evidence. Furthermore, our model results support the idea that such a northern lowlands ocean had to be sufficiently large to sustain the hydrologic cycle in a warm early climate [e.g., *Parker et al*., 1993; *Di Achille and Hynek,* 2010; *Luo et al*., 2017; *Citron et al.,* 2018; *Duran et al.,* 2019].

The *von Paris et al.* [2015] study used a 1-D model to compute runoff rates in cold $CO_2$-$H_2O$ climates and found (quite expectedly) that they are orders of magnitude lower than what has been inferred for valley network terrains. On this basis, they concluded that the early climate must have been cold and icy. However, such a conclusion avoids explaining why the valley networks exist to begin with. Instead, we argue that the reason why some climate models fail to warm the climate [e.g., *von Paris et al.*, 2015; *Wordsworth et al.,* 2013] stems from assuming an initially cold and icy climate that undergoes transient warming episodes. Our results do not support the idea that exotic transient warming mechanisms in icy climates were necessary to warm early Mars. The challenges with these mechanisms have been previously discussed [e.g., *Ramirez* 2017; *Ramirez and Craddock,* 2018]. As we have also argued in this work, it is unlikely that such events would have produced sufficient runoff.

Indeed, if the valleys were formed in < ~1 -10 million years, such transient warming episodes may have been extraneous. Alternatively, a single brief warm and semi-arid episode lasting a hundred thousand to a few million years on a volcanically-active early Mars, one with seasonally higher precipitation, could have generated the necessary runoff and explain the observed erosion [*Ramirez and Craddock*, 2018]. At colder temperatures, the ocean freezes and surface erosion would become inadequate to produce the observed fluvial features. This warm climate may have been facilitated by a sufficiently potent $CO_2$-$H_2$ greenhouse [*Ramirez et al.* 2014a; *Ramirez,* 2017]. This does not necessarily mean that Mars was warm during the entire Noachian (though it may have been), but at the very least, conditions should have been warm during the late Noachian-early Hesperian climate optimum. Revised geologic mapping reveals that peak tectonic deformation within Tharsis had occurred simultaneously with valley network formation [e.g., *Bouley et al.,* 2018], further indicative of a growing Tharsis bulge on a volcanically active early Mars.

Future work requires better estimates of $CO_2$-$H_2$ collision-induced absorption, improving estimates of the required atmospheric pressures and concentrations on early Mars. We also stress that the length of the early warm and wet period remains highly uncertain and depends on current geologic estimates of erosion rates. Updated erosion rate estimates would improve estimates of the duration of the climate optimum during valley formation. Continued work aimed at improving paleopressure estimates would also be very instructive.

**Acknowledgements:** All the data and programs used to produce all figures and tables in this work are available at https://doi.org/10.5281/zenodo.3550244. R.M.R. acknowledges funding from the Earth-Life Science Institute (ELSI) and also from the Astrobiology Center grant proposal grant number JY310064. R.A.C. acknowledges support from NASA MDAP grant 80NSSC17K0454.  T.U. acknowledges support from JSPS Grants-in-Aid for Scientific Research (19H01960, 17H06459, 15KK0153, 16H04073).   Valley network data used are listed in Tables 1 and 2. We also greatly appreciated comments and suggestions from Alejandro Soto, Michael Mischna and an anonymous reviewer.

Batalha, N.E., et al (2016), Climate cycling on early Mars caused by the carbonate–silicate cycle, *Earth and Planetary Science Letters* 455, 7-13.

Bin, Jiayu, Feng Tian, and Lei Liu (2018), New inner boundaries of the habitable zones around M dwarfs, *Earth and Planetary Science Letters* 492, 121-129.

Bishop, J.L., et al. (2018), Surface clay formation during short-term warmer and wetter conditions on a largely cold ancient Mars, *Nature Astronomy* 2,3, 206.

Bouquety, A., et al. (2019), Morphometric evidence of 3.6 Ga glacial valleys and glacial cirques in martian highlands: South of Terra Sabaea, *Geomorphology* 334, 91-111.

Bouley, S., et al. (2016). Late Tharsis formation and implications for early Mars. *Nature*, *531*, 7594, p.344.

Bouley, Sylvain, et al. (2018), The revised tectonic history of Tharsis, *Earth and Planetary Science Letters* 488, 126-133.

Burch, Darrell E., et al. (1969), Absorption of infrared radiant energy by $CO_2$ and $H_2O$. IV. Shapes of collision-broadened $CO_2$ lines, *JOSA* 59, 3, 267-280.

Caballero, Rodrigo, and Peter L. Langen (2005), The dynamic range of poleward energy transport in an atmospheric general circulation model, *Geophysical Research Letters* 32, 2.

Caldeira, Ken, and James F. Kasting (1992), Susceptibility of the early Earth to irreversible glaciation caused by carbon dioxide clouds, *Nature* 359, 6392, 226.

Cataldo, Joseph C., et al. (2010), Prediction of transmission losses in ephemeral streams, Western USA, *The Open Hydrology Journal* 4, 1.

Cawley, J. C., & Irwin, R. P. (2018), Evolution of escarpments, pediments, and plains in the Noachian highlands of Mars, *Journal of Geophysical Research: Planets*, 123, 3167–3187.

Citron, Robert I., Michael Manga, and Douglas J. Hemingway (2018),Timing of oceans on Mars from shoreline deformation, *Nature* 555, 7698, 643.

Craddock, R. A., T. A. Maxwell, and A. D. Howard (1997), Crater morphometry and modification in the Sinus Sabaeus and Margaritifer Sinus regions of Mars, *Journal of Geophysical Research*, 102, 13,321– 13,340.

Craddock, R.A., and A.D. Howard (2002), The case for rainfall on a warm, wet early Mars, *Journal of Geophysical Research: Planets* 107, E11, 21-1.

Craddock, R.A., and R. Greeley (2009), Minimum estimates of the amount and timing of gases released into the martian atmosphere from volcanic eruptions, *Icarus* 204, 2, 512-526.
50

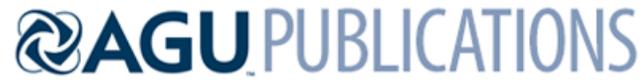

**CLIMATE SIMULATIONS OF EARLY MARS WITH ESTIMATED PRECIPITATION, RUNOFF, AND EROSION RATES**

Ramses M. Ramirez[1,2], Robert A. Craddock[3] and Tomohiro Usui[1,4]

[1]Earth-Life Science Institute, Tokyo Institute of Technology, Tokyo, Japan

[2]Space Science Institute, Boulder, Co, USA

[3]Center for Earth and Planetary Studies, National Air and Space Museum, Smithsonian Institution, Washington D.C., USA

[4]Institute of Space and Astronautical Science (ISAS), Japan Aerospace Exploration Agency, Tokyo, Japan

The following are additional sensitivity studies on special scenarios that complement the results in the main text.

*S1: Impact of clouds on surface temperature, precipitation, and runoff rates*

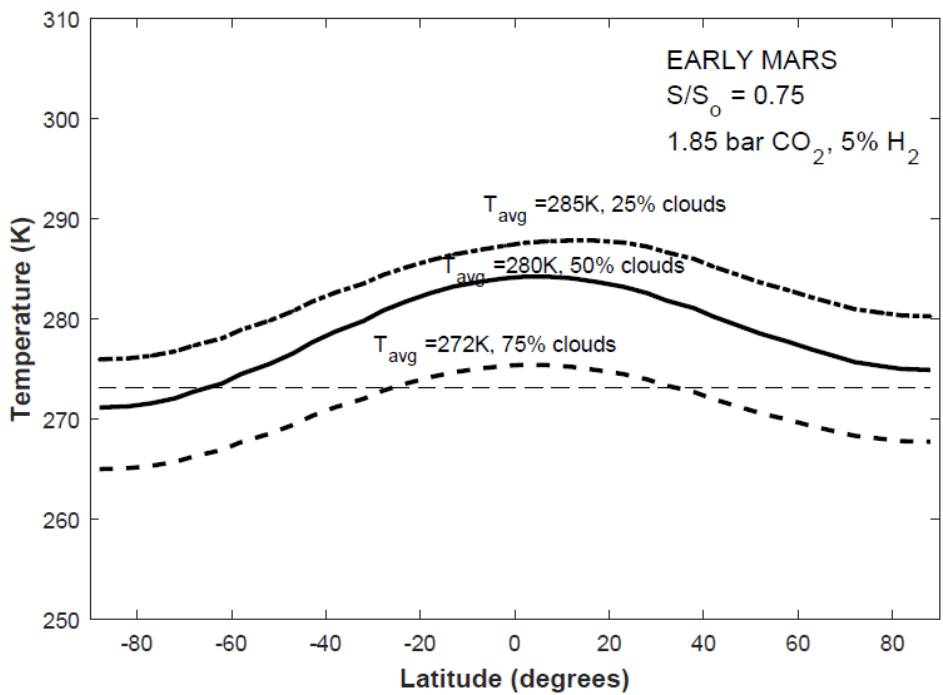

**Figure S1:** Planetary surface temperature response to cloud cover fractions of 25% (dashed-dotted line), 50% (solid line), and 75% (dashed line), respectively, for the 1.85 bar $CO_2$, 5% $H_2$ atmosphere with large ocean.

Although we have assumed an Earth-like 50% cloud cover for most of our analysis, we have evaluated the effect of higher (75%) and lower (25%) cloud cover on our main results, again assuming a constant relative humidity (Figures S1 – S3). For the 1.85 bar $CO_2$ 5% $H_2$ case, the mean surface temperature increases from ~280 K to 285 K at 25% cloud cover and decreases to 272 K at 75% cloud cover, respectively (Figure S1).

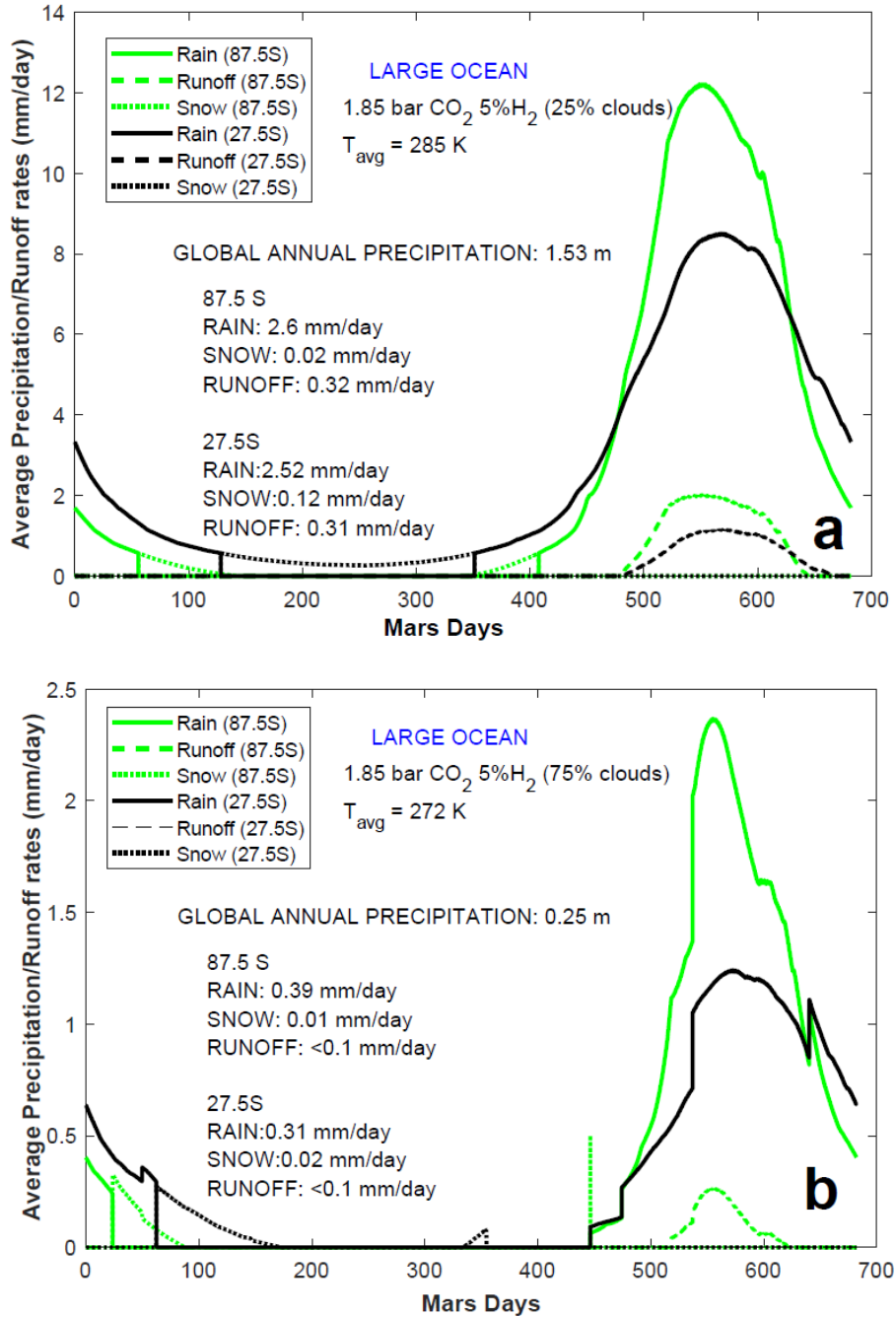

**Figure S2:** Precipitation and runoff rates during the Martian year at (green) 87.5°S and (black) 27.5°S latitudes for a 1.85 bar $CO_2$ 5% $H_2$ atmosphere with (a) 25% and (b) 75% cloud cover and a large ocean, respectively. Same color scheme as in Figures 8 – 11.

Precipitation rates are especially responsive to the surface temperature changes arising from cloud cover variations. From 50% to 25% cloud cover, the global annual precipitation rates increase from ~1.2m to ~1.5m whereas they decrease to 0.25m at 75% cloud cover (Figure 8a; S2ab). Minor snow is produced in both scenarios although mean runoff rates are much higher (~0.3 mm/day vs. <0.1 mm/day) in the low cloud cover case at corresponding latitudes (Figure S2).

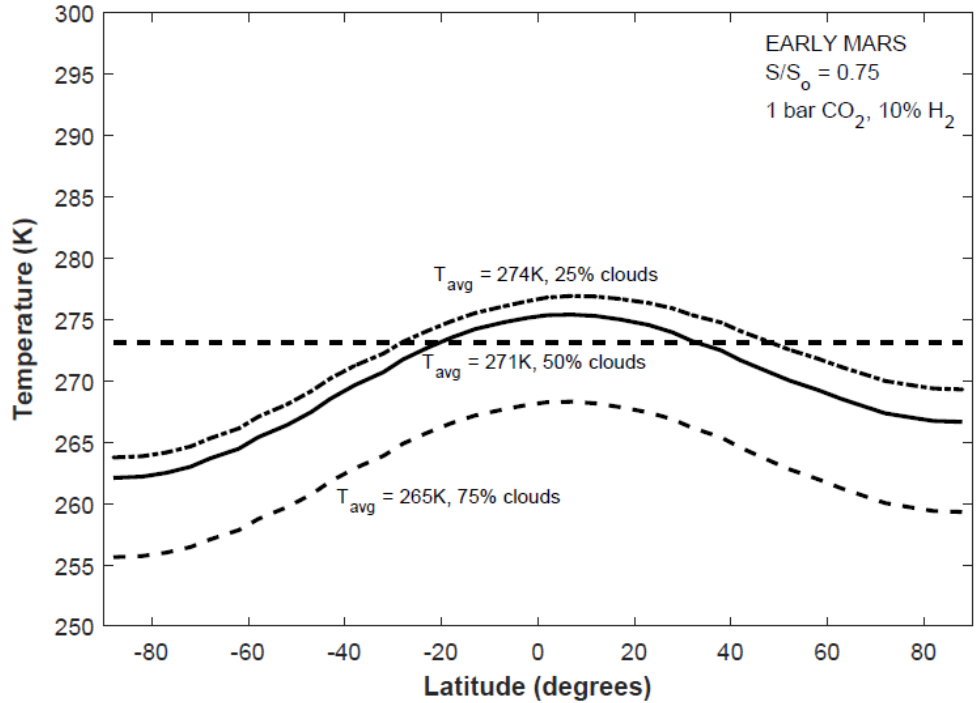

**Figure S3:** Planetary surface temperature response to cloud cover fractions of 25% (dashed-dotted line), 50% (solid line), and 75% (dashed line), respectively, for the 1 bar $CO_2$ 10% $H_2$ atmosphere.

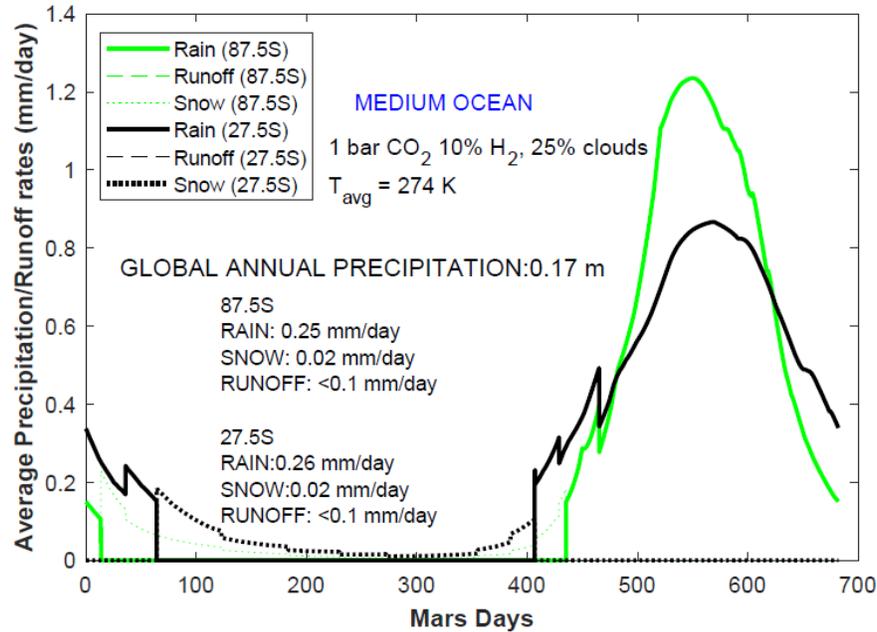

**Figure S4:** Precipitation and runoff rates during the Martian year at (green) 87.5°S and (black) 27.5°S latitudes for a 1 bar $CO_2$ 10% $H_2$ atmosphere with 25% cloud cover and a medium-sized ocean. Same color scheme as in Figures 8 – 11, S2.

We performed a similar set of sensitivity studies for the 1 bar $CO_2$ 10% $H_2$ case with the medium-sized ocean (Figures S3 – S4). From the baseline 50% cloud cover case, mean surface temperatures increased from 271 K to 274 K for low cloud cover and decreased from 271 K to 265K for high cloud cover, respectively (Figure 9b; Figure S3). The surface temperatures at 75% cloud cover were low enough for the hydrologic cycle to shut down, leading to a planet with no appreciable surface runoff. Although some rainfall occurred at low cloud cover (~0.25 mm/year; Figure S4), it was not enough to produce more than a negligible amount of snowfall or runoff (Figure S4), strengthening the hypothesis that a relatively large ocean in the late Noachian-early Hesperian, like that of *DiAchille and Hynek* (2010), more easily explains formation of valley networks and other major fluvial features.

In reality, if surface relative humidity becomes higher in warmer climates [e.g., *Ramirez et al.* 2014b; *Kasting et al.* 2014], temperatures and precipitation rates for low cloud cover may be somewhat higher than calculated here. The inverse may be true at higher cloud cover.

*S2: Diffusion coefficient sensitivity study for warm and very cold early Mars climates*

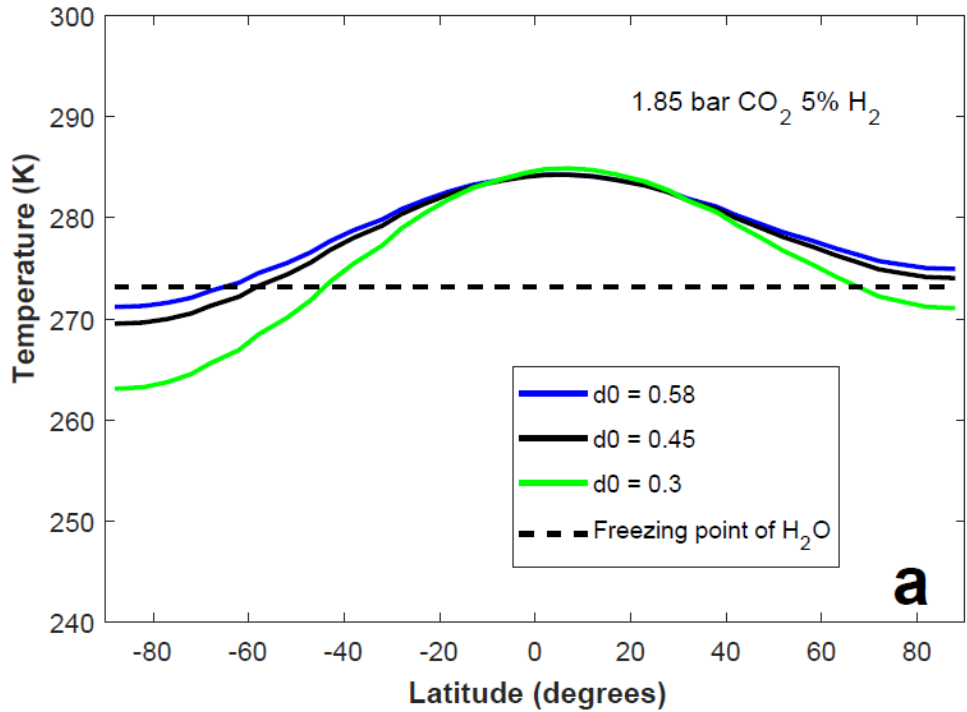

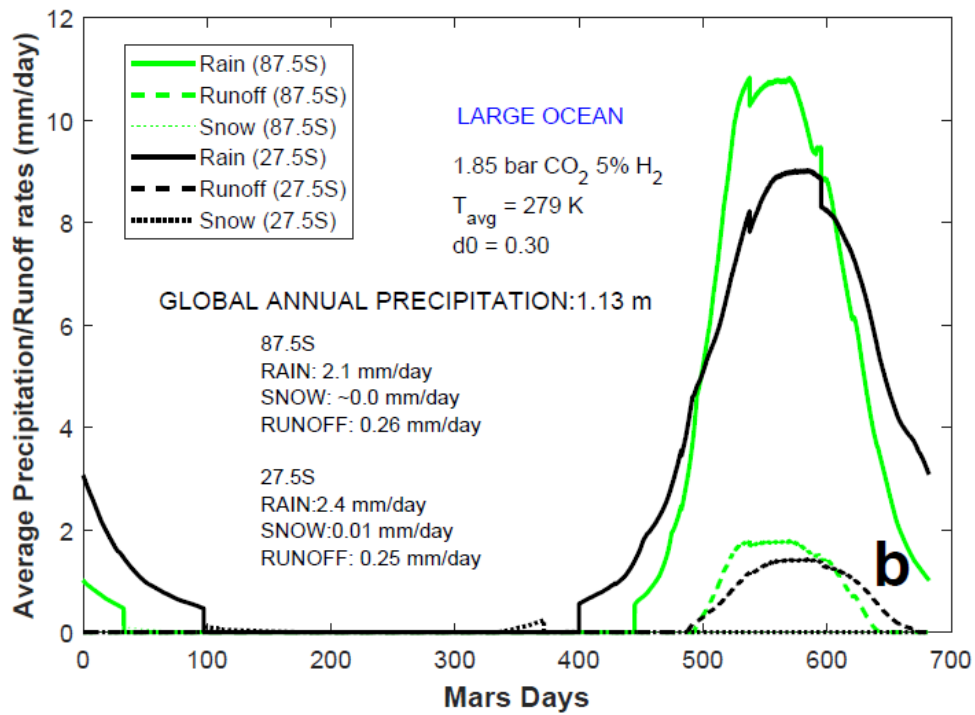

**Figure S5:** A comparison of (a) latitudinal surface temperatures for different $D_o$ values and (b) precipitation and runoff rates at $D_o = 0.3$ for the 1.85 bar $CO_2$ 5% $H_2$ case. The precipitation/runoff plot uses the same color scheme as in Figures 8 – 11, S2, S4.

This sensitivity study assesses the scenario that a warm early Mars may have had a somewhat lower heat transport efficiency than the Earth. Our main trends are rather robust to significant decreases in $D_o$. Planetary surface temperatures change by no more than 1 or 2 degrees for a $D_o$ decrease from 0.58 to 0.45 (~22%) (Figure S5a), yielding global precipitation patterns that are very similar to those for the baseline case (Figure 8a). Although polar temperatures are a few degrees colder at $D_o = 0.3$ (Figure S5a), global precipitation rates are still ~95% that of the baseline case (Figure S5b; Figure 8a). Rainfall rates at the southern latitude band are also very similar to those for the baseline case (Figure S5b; Figure 8a), illustrating that transport efficiency is still high enough to maintain warm temperatures and high rainfall throughout the planet. Runoff rates (~0.25 mm/day), are similar to the baseline case as well. We consider even lower $D_o$ values to be unrealistic for a warm early Mars because that would suggest comparable latitudinal gradients to Earth's. However, at the higher atmospheric pressures considered here, latitudinal gradients should be reduced, following equation 9.

We also assess whether changes in $D_o$ could affect the latitudinal temperature distribution of cold climates on early Mars. We find that decreasing $D_o$ has nearly no impact on the latitudinal temperature distribution of our 0% $H_2$ case because heat transport is too small at such low mean surface temperatures (< 210 K)( Figure 5a) for effective transport to occur. For the same reason, even a reduction in surface albedo does not appreciably increase surface temperatures in this case.